\documentclass[twocolumn,resetfootnote]{aastex701}

\usepackage[T1]{fontenc}
\usepackage[version=4]{mhchem}
\usepackage{booktabs}
\usepackage{amsmath}
\usepackage{xspace}

\newcommand{\PROTEUS}{\texttt{PROTEUS}\xspace}
\newcommand{\Aragog}{\texttt{Aragog}\xspace}
\newcommand{\Zalmoxis}{\texttt{Zalmoxis}\xspace}
\newcommand{\CALLIOPE}{\texttt{CALLIOPE}\xspace}
\newcommand{\PALEOS}{\texttt{PALEOS}\xspace}
\newcommand{\atmodeller}{\texttt{atmodeller}\xspace}
\newcommand{\AGNI}{\texttt{AGNI}\xspace}
\newcommand{\Mearth}{M_\mathrm{Earth}}
\newcommand{\Rearth}{R_\mathrm{Earth}}
\newcommand{\fO}{f_\mathrm{O_2}}

\newcommand{\zdoi}[1]{\href{https://doi.org/10.5281/zenodo.#1}{doi:10.5281/\allowbreak zenodo.\allowbreak#1}}

\shorttitle{Super-Earth Interiors Shrink by About 10\% as They Crystallise}
\shortauthors{Lichtenberg, Attia, Nicholls, Sastre, et al.}

\graphicspath{{./}{figures/}}

\begin{document}

\title{Super-Earth Interiors Shrink by About 10\% as They Crystallise}

\author[0000-0002-3286-7683]{Tim Lichtenberg}
\email[show]{tim.lichtenberg@rug.nl}
\affiliation{Kapteyn Astronomical Institute, University of Groningen, P.O.\ Box 800, 9700 AV Groningen, The Netherlands}

\author[0000-0002-7971-7439]{Mara Attia}
\email{}
\affiliation{Kapteyn Astronomical Institute, University of Groningen, P.O.\ Box 800, 9700 AV Groningen, The Netherlands}

\author[0000-0002-8368-4641]{Harrison Nicholls}
\email{}
\affiliation{Institute of Astronomy, University of Cambridge, Madingley Road, Cambridge CB3 0HA, UK}

\author[0009-0008-7799-7976]{Mariana Sastre}
\email{}
\affiliation{Kapteyn Astronomical Institute, University of Groningen, P.O.\ Box 800, 9700 AV Groningen, The Netherlands}

\author[0000-0002-0673-4860]{Dan J.\ Bower}
\email{}
\affiliation{Department of Earth and Planetary Sciences, ETH Zurich, Zurich, Switzerland}

\author[0009-0000-6847-4331]{Karen Stuitje}
\email{}
\affiliation{Kapteyn Astronomical Institute, University of Groningen, P.O.\ Box 800, 9700 AV Groningen, The Netherlands}

\author[0009-0007-4663-1456]{Flavia C.\ Pascal}
\email{}
\affiliation{Kapteyn Astronomical Institute, University of Groningen, P.O.\ Box 800, 9700 AV Groningen, The Netherlands}
\affiliation{Institute of Astronomy, University of Cambridge, Madingley Road, Cambridge CB3 0HA, UK}

\author[0000-0002-5422-8794]{Laurent Soucasse}
\email{}
\affiliation{IMEC, Kapeldreef 75, 3001 Leuven, Belgium}
\affiliation{Netherlands eScience Center, Science Park 402, 1098 XH Amsterdam, The Netherlands}

\author[0000-0003-3714-5855]{Daniel Apai}
\email{}
\affiliation{Department of Astronomy/Steward Observatory, University of Arizona, Tucson, AZ 85721, USA}
\affiliation{Lunar and Planetary Laboratory, University of Arizona, Tucson, AZ 85721, USA}

\author[0000-0002-6033-960X]{Patrick Bos}
\email{}
\affiliation{Kapteyn Astronomical Institute, University of Groningen, P.O.\ Box 800, 9700 AV Groningen, The Netherlands}
\affiliation{Center for Information Technology, University of Groningen, P.O.\ Box 11044, 9700 CA Groningen, The Netherlands}

\author[0009-0002-9247-2437]{Robb Calder}
\email{}
\affiliation{Institute of Astronomy, University of Cambridge, Madingley Road, Cambridge CB3 0HA, UK}

\author[0009-0009-7228-7809]{Lorenzo Cesario}
\email{}
\affiliation{Kapteyn Astronomical Institute, University of Groningen, P.O.\ Box 800, 9700 AV Groningen, The Netherlands}

\author[0000-0003-4987-6591]{Lisa Dang}
\email{}
\affiliation{Department of Physics and Astronomy and Waterloo Centre for Astrophysics, University of Waterloo, Waterloo, ON N2L 3G1, Canada}

\author[0009-0008-3326-9715]{Emeline Decocq}
\email{}
\affiliation{Kapteyn Astronomical Institute, University of Groningen, P.O.\ Box 800, 9700 AV Groningen, The Netherlands}

\author[0009-0005-6677-1929]{Marijn van Dijk}
\email{}
\affiliation{Kapteyn Astronomical Institute, University of Groningen, P.O.\ Box 800, 9700 AV Groningen, The Netherlands}

\author[0000-0001-7864-6627]{Mohammad Farhat}
\email{}
\affiliation{Department of Astronomy, University of California, Berkeley, Berkeley, CA 94720-3411, USA}
\affiliation{Department of Earth and Planetary Science, University of California, Berkeley, Berkeley, CA 94720-4767, USA}

\author[0000-0003-4815-2874]{Kaustubh Hakim}
\email{}
\affiliation{Royal Observatory of Belgium, Avenue Circulaire 3, 1180 Brussels, Belgium}
\affiliation{Institute of Astronomy, KU Leuven, Celestijnenlaan 200D, 3001 Leuven, Belgium}

\author[0000-0001-8477-2523]{Tadahiro Kimura}
\email{}
\affiliation{UTokyo Organization for Planetary Space Science (UTOPS), University of Tokyo, Hongo, Bunkyo-ku, Tokyo 113-0033, Japan}
\affiliation{Kapteyn Astronomical Institute, University of Groningen, P.O.\ Box 800, 9700 AV Groningen, The Netherlands}

\author[0009-0009-7323-6755]{Imre Kisv\'{a}rdai}
\email{}
\affiliation{Kapteyn Astronomical Institute, University of Groningen, P.O.\ Box 800, 9700 AV Groningen, The Netherlands}

\author[0000-0002-3291-6887]{Sebastiaan Krijt}
\email{}
\affiliation{Department of Physics and Astronomy, University of Exeter, Stocker Road, Exeter EX4 4QL, UK}

\author[0000-0002-0747-8862]{Yamila Miguel}
\email{}
\affiliation{Leiden Observatory, Leiden University, Einsteinweg 55, 2333 CC Leiden, The Netherlands}
\affiliation{SRON Netherlands Institute for Space Research, Niels Bohrweg 4, 2333 CA Leiden, The Netherlands}

\author[0009-0005-9147-0431]{Ioannis Panagiotou}
\email{}
\affiliation{Kapteyn Astronomical Institute, University of Groningen, P.O.\ Box 800, 9700 AV Groningen, The Netherlands}

\author[0009-0009-5036-3049]{Emma Postolec}
\email{}
\affiliation{Kapteyn Astronomical Institute, University of Groningen, P.O.\ Box 800, 9700 AV Groningen, The Netherlands}

\author[0000-0001-8355-2107]{Martin Schlecker}
\email{}
\affiliation{European Southern Observatory, Karl-Schwarzschild-Str.\ 2, 85748 Garching, Germany}

\author[0000-0002-6892-6948]{Sara Seager}
\email{}
\affiliation{Department of Earth, Atmospheric and Planetary Sciences, Massachusetts Institute of Technology, Cambridge, MA 02139, USA}

\author[0000-0002-0794-2717]{Anat Shahar}
\email{}
\affiliation{Earth and Planets Laboratory, Carnegie Institution for Science, 5241 Broad Branch Road NW, Washington, DC 20015, USA}

\author[0000-0002-8713-1446]{Oliver Shorttle}
\email{}
\affiliation{Institute of Astronomy, University of Cambridge, Madingley Road, Cambridge CB3 0HA, UK}
\affiliation{Department of Earth Sciences, University of Cambridge, Downing Street, Cambridge CB2 3EQ, UK}

\author[0000-0002-1462-1882]{Paolo A.\ Sossi}
\email{}
\affiliation{Department of Earth and Planetary Sciences, ETH Zurich, Zurich, Switzerland}

\author[0000-0001-5828-8885]{Wim van Westrenen}
\email{}
\affiliation{Department of Earth Sciences, Vrije Universiteit Amsterdam, De Boelelaan 1100, 1081 HZ Amsterdam, The Netherlands}

\begin{abstract}
Super-Earth exoplanets are among the most abundant planets known, yet their bulk densities leave the interior state degenerate.
The static structure models used to interpret them, and the interior retrievals built on them, typically describe the cold, solidified end state of an evolution that begins hot and molten.
During this magma ocean stage the interior, the outgassed atmosphere, and the surface co-evolve and set the long-term climate and geophysics of super-Earths.
We develop and validate a fully coupled model for the structural and thermal evolution of super-Earth exoplanets within the \PROTEUS framework, including new and upgraded models of the interior structure, mantle energetics, and volatile outgassing.
In volatile-poor super-Earths of $1$ to $10$ Earth masses, the silicate interior contracts by about $10\,\%$ of its molten radius through cooling and crystallisation, nearly independent of planet mass and driven by the thinning silicate shell alone.
The solidified radius is set by planetary mass and core fraction, insensitive to the host star, irradiation, and initial thermal state.
In contrast, volatile-rich super-Earths at and above about $5$ Earth masses may not solidify: their thick outgassed atmospheres throttle the surface heat loss until the interior settles into a deep magma ocean, keeping the planet inflated and limiting the contraction to about half its volatile-poor value.
Mantle contraction alone thus shapes the low-mass exoplanet transit population, motivating joint interpretation of atmospheric and geophysical signatures in upcoming exoplanet surveys.
\end{abstract}


\section{Introduction}
\label{sec:intro}

Super-Earths, rocky planets of roughly one to ten Earth masses, are among the most abundant planets known, yet their interiors remain among the least constrained \citep{Lichtenberg2025TrGeo751L}.
More precisely, we here define super-Earths as planets with rocky interiors, a class set by interior composition rather than presence or thickness of an atmosphere.
Core size, mantle composition, and a volatile envelope trade off against one another, so a measured mass and radius permit many internal structures at the same bulk density \citep{Valencia2006Icar181545V,Seager2007ApJ6691279S,Dorn2015AA577A83D,Dorn2017AA597A37D,Unterborn2018NatAs2297U,Steffen2025ApJ991L33S}.
The radius gap that separates rocky super-Earths from sub-Neptunes, centred near $1.8\,\Rearth$ and spanning roughly $1.5$ to $2.0\,\Rearth$ \citep{Fulton2017AJ154109F,VanEylen2018MNRAS4794786V,Owen2017ApJ84729O}, makes this degeneracy harder to resolve.
A radius alone cannot be used to infer whether a planet is a bare rock, a volatile-stripped core, or a water-rich world \citep{Zeng2019PNAS1169723Z,Luque2022Sci3771211L}.
Interpretation rests on vertically resolved interior-structure models, which usually compute the radius of a cold, fully solidified planet at a prescribed composition \citep{Seager2007ApJ6691279S,Zeng2016ApJ819127Z,Dorn2017AA597A37D,Baumeister2025SSRv221123B}, the static end state of an evolution whose hotter phases are not part of the calculation.

That end state is reached only after an initial molten stage.
A rocky planet emerges from accretion hot enough to melt its mantle, and the resulting magma ocean partitions volatiles between interior and atmosphere, sets the surface oxidation state, and paces the cooling and crystallisation \citep{ElkinsTanton2012AREPS40113E,Lebrun2013JGRE1181155L,Hamano2013Natur497607H,Schaefer2016ApJ82963S,Bower2019AA631A103B,Lichtenberg2023ASPC534907L}, with the outgassed atmosphere controlling the radiative cooling and its composition set by the oxidation state of the melt \citep{Sossi2023EPSL60117894S,Bower2022PSJ393B}.
This coupling of interior, surface, and atmosphere \citep{Lichtenberg2025Sci390S3660L} has been modelled in detail at Earth mass \citep{ElkinsTanton2008EPSL271181E,Lebrun2013JGRE1181155L,Bower2018PEPI27449B,Nicholls2024JGRE12908576N}.
Such calculations are evolutionary rather than static: they advance the interior and atmosphere forward in time, and the solidified structures that static models solve for are their end states.
To date, where such evolutionary coupling has been extended to super-Earth masses, the interior structure is held static in time and only its thermal and chemical states are evolved \citep[e.g.][]{Schaefer2016ApJ82963S,KrissansenTotton2022ApJ933115K,Nicholls2026NatAs10809N}.
What has not been treated is the joint evolution of the interior structure itself together with the volatile outgassing and the climate.
At super-Earth masses this structural coupling is essential, because higher pressures change the equation of state, the melting curve, and the volatile solubilities together \citep{Boley2023ApJ954202B,Baumeister2025SSRv221123B}.

Two simplifications inherited from the Earth-mass case become consequential at super-Earth masses.
First, the interior structure is typically held fixed while the mantle cools, so the melt-to-solid contraction of the radius is neither captured nor its mass dependence known.
This contraction, first indicated for an evolving terrestrial mantle by \citet{Bower2019AA631A103B}, is amplified by the larger thermal expansivity of silicate liquids relative to their solids \citep{Wolf2018PEPI27859W,Stixrude2009EPSL278226S} and modulated at high pressure by the compression state of the interior \citep{Luo2025JGRE13008678L}.
Second, the oxidation state of the melt, how oxidised the mantle is, conventionally referenced to the iron-w\"ustite (Fe-FeO) mineral equilibrium, is almost always imposed as a fixed offset held constant through the evolution \citep{Sossi2023EPSL60117894S,Bower2022PSJ393B,Nicholls2024JGRE12908576N}.
Other coupled treatments instead prescribe a fixed volatile budget and compute the outgassing at equilibrium \citep{Schaefer2016ApJ82963S,KrissansenTotton2022ApJ933115K}, so how volatiles and their oxidation state pass between interior and atmosphere differs from model to model.
Treating the planetary oxygen content as a conserved elemental inventory lets the surface fugacity evolve as a derived property instead; the difference between this free-oxygen accounting and the fixed-fugacity treatment has not yet been quantified over a coupled grid.
That the available treatments diverge on exactly these choices motivated a recent community intercomparison of coupled magma ocean models \citep{Lichtenberg2026PSJ7108L,Nicholls2026arXiv260624757N}, and coupled analyses of individual super-Earths show how strongly the treatment shapes the outcome \citep{Nicholls2026NatAs10809N}.

A fully coupled treatment is now tractable: multiphase equations of state to super-Earth pressures \citep{Wolf2018PEPI27859W,Attia2026subm}, entropy-based mantle evolution across phase boundaries \citep{Bower2018PEPI27449B}, and thermochemical outgassing across the C-H-O-N-S system \citep{Bower2025ApJ99559B,Hakim2026MNRAS546ag133H} can be solved together at every step \citep{Lichtenberg2021JGRE12606711L,Nicholls2024JGRE12908576N}.
These choices interact through the shared equation of state, opacity, and speciation, so their consequences emerge only when the full framework is closed and integrated forward in time.

This paper has two aims: to establish a validated, fully coupled framework for the evolution of differentiated super-Earths, and to determine how a rocky planet's radius changes as it crystallises.
Within the modular \PROTEUS\footnote{\url{https://proteus-framework.org}} framework \citep{Lichtenberg2021JGRE12606711L,Nicholls2024JGRE12908576N} we introduce three new modules (Section~\ref{sec:methods}): \Zalmoxis, which solves the interior structure on the phase-aware \PALEOS\ equation of state of \citet{Attia2026subm} to the multi-TPa pressures of super-Earth interiors; \Aragog, which evolves the mantle energetics as a direct descendant of the SPIDER code \citep{Bower2018PEPI27449B}; and \atmodeller\ \citep{Bower2025ApJ99559B}, which computes the volatile outgassing.
With these modules in place we follow the evolution of super-Earths and its effect on their radius.
We vary planet mass over $1$ to $10\,\Mearth$, the oxidation state over a broad range around the iron-w\"ustite buffer in two redox treatments, the volatile inventory across four prescriptions, and the host star, irradiation, core-mass fraction, and initial thermal state.
We find that the silicate mantle of super-Earths contracts measurably as it crystallises (Section~\ref{sec:results}), by about $10\,\%$ of its molten radius, nearly independent of planet mass and insensitive to the host star, with the solidified radius independent of the initial state.

The remainder of the paper develops this result.
Section~\ref{sec:methods} presents the coupled framework and its new interior-structure, mantle-energetics, and outgassing modules.
Section~\ref{sec:framework_validation} validates them against analytic limits, published structure models, and the community magma ocean intercomparison.
Section~\ref{sec:results} reports the contraction and how it responds to planet mass, oxidation state, volatile inventory, and stellar environment.
Section~\ref{sec:discussion} draws out the consequences for interpreting super-Earth radii.
Section~\ref{sec:conclusions} concludes, and Appendix~\ref{sec:validation} collects the full validation suite.

\section{Methods}
\label{sec:methods}

\subsection{Coupled interior-atmosphere framework}
\label{sec:methods:proteus}

\begin{figure}
\centering
\includegraphics[width=\columnwidth]{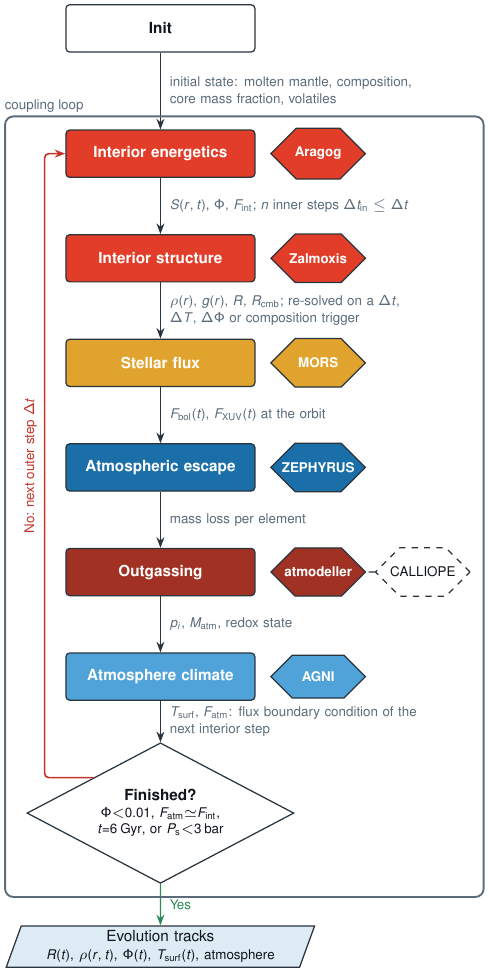}
\caption{The \PROTEUS\ coupling loop for super-Earth evolution.
Each outer time step advances the coupled interior, surface, and atmosphere in turn: the interior energetics (\Aragog), interior structure (\Zalmoxis), stellar flux (\texttt{MORS}), atmospheric escape (\texttt{ZEPHYRUS}), volatile outgassing (\atmodeller, with \CALLIOPE\ retained as a cross-validation reference, shown dashed), and atmosphere climate (\AGNI) appear as the coloured process boxes, each with its module at right, evaluated in turn and passing the quantities labelled on the arrows; the interior structure is re-solved only when a threshold is crossed (Section~\ref{sec:methods:zalmoxis}), not at every step.
The surface temperature and the atmospheric flux $F_\mathrm{atm}$ set the boundary condition of the next interior step.
The step repeats until the mantle solidifies (melt fraction $\Phi < 0.01$), the interior reaches a radiative quasi-steady state in which the net energy imbalance falls below $1\,\mathrm{W\,m^{-2}}$ ($F_\mathrm{atm} \simeq F_\mathrm{int}$; Section~\ref{sec:results:volatiles}), the age reaches $6\,\mathrm{Gyr}$, or the atmosphere desiccates ($P_\mathrm{s} < 3\,\mathrm{bar}$), whereupon the run returns the evolution tracks.
\label{fig:proteus_loop}}
\end{figure}

We simulate the coupled structural and thermal evolution of rocky super-Earths using the modular \PROTEUS\ framework \citep{Lichtenberg2021JGRE12606711L,Nicholls2024JGRE12908576N,Nicholls2025MNRAS5362957N,Nicholls2025MNRAS5412566N,Nicholls2026NatAs10809N,vanDijk2026PSJ794V,Lichtenberg2026PSJ7108L,Calder2026MNRAS549g1007C,Sastre2026arXiv260620249S,Postolec2026arXiv260715011P,Postolec2026arXiv260903144P,Cesario2026arXiv260713793C,Panagiotou2026arXiv260715204P}.
Starting from a fully molten initial state, as expected after the energetic final stages of rocky planet accretion characterised by giant impacts \citep{Solomatov2007eveabook91S,ElkinsTanton2012AREPS40113E,Lichtenberg2023ASPC534907L}, \PROTEUS\ time-steps the planet by sharing the evolving planetary state (e.g.\ surface temperature, melt fraction, atmospheric composition, radiation fluxes, planetary radius) across a set of submodules at every outer time step (Figure~\ref{fig:proteus_loop}).
The planet advances on an outer time step $\Delta t$, over which the volatile budgets, stellar flux, and atmospheric state are held fixed while the interior energetics are sub-cycled on $n$ inner steps $\Delta t_\mathrm{in} \le \Delta t$; the interior structure is re-solved only when a time, temperature, melt-fraction, or composition threshold is crossed (Section~\ref{sec:methods:zalmoxis}) rather than at every step.
The framework adopts a four-domain decomposition.
The \emph{interior structure} of the planet (mass, radius, density profile, gravity, and core-mantle boundary state) is solved by \Zalmoxis, a new interior-structure module developed for this work.
The \emph{interior energetics} (mantle temperature profile, surface heat flux, melt fraction) is solved by \Aragog.
The \emph{atmosphere} (radiative-convective temperature profile and outgoing radiative flux) is solved by \AGNI\ \citep{Nicholls2025JOSS107726N,Nicholls2025MNRAS5362957N}.
The \emph{stellar} domain (the time-evolving bolometric and X-ray-to-ultraviolet flux at the planet) is provided by the \texttt{MORS} stellar evolution model \citep{Johnstone2021AA649A96J,Spada2013ApJ77687S}.
A separate \emph{outgassing} step (\atmodeller) solves for the partitioning of C-H-O-N-S volatiles between the atmosphere and the magma ocean at thermochemical-solubility equilibrium at the surface boundary \citep{Bower2025ApJ99559B,Hakim2026MNRAS546ag133H}.
A further \emph{escape} step removes volatiles from the top of the atmosphere by XUV-driven hydrodynamic escape, computed by the \texttt{ZEPHYRUS} module \citep{Postolec2026arXiv260715011P}: the mass-loss rate follows an energy-limited formulation with the Roche-lobe correction of \citet{Erkaev2007AA472329E}, driven by the X-ray-to-ultraviolet flux delivered by \texttt{MORS}, and the escaped mass is deducted from the elemental volatile inventory at each outer step before the next outgassing solve.
At each outer step the interior thermal evolution is subjected to a flux boundary condition derived from the atmospheric solution, ensuring a closed energy balance between the molten mantle and the radiating top of the atmosphere \citep{Nicholls2024JGRE12908576N}; the formulation is described in Section~\ref{sec:methods:aragog:bc}, and a grey-body surface boundary condition is used only in the standalone conservation tests of Appendix~\ref{sec:app:aragog}.

This work introduces four substantive updates to the framework relative to its earlier published applications: the new \Zalmoxis\ interior structure solver, validated here up to $20\,\Mearth$ against published structure models (Section~\ref{sec:methods:zalmoxis}); the new \Aragog\ interior energetics solver, which supersedes the SPIDER mantle thermal evolution code \citep{Bower2018PEPI27449B,Bower2019AA631A103B,Bower2022PSJ393B} for all coupled runs in this work (Section~\ref{sec:methods:aragog}); the new \atmodeller\ outgassing solver \citep{Bower2025ApJ99559B,Hakim2026MNRAS546ag133H}, which replaces the \CALLIOPE\ submodule used in earlier work (Section~\ref{sec:methods:atmodeller}); and an oxygen-conserving outgassing mode, implemented in both \atmodeller\ and \CALLIOPE, in which the planetary oxygen budget is prescribed as a conserved elemental inventory, enabling the surface oxygen fugacity to become a property derived from the equilibrium rather than an imposed boundary condition (Section~\ref{sec:methods:atmodeller}).
SPIDER and \CALLIOPE\ are retained within \PROTEUS\ as cross-validation references for the new modules; the parity comparisons are reported in Appendices~\ref{sec:val:aragog} and~\ref{sec:app:calliope_atmodeller}.
Each simulation is initialised in a fully molten state and terminates at whichever comes first: solidification of the global mantle ($\Phi < 0.01$), a radiative quasi-steady state, a maximum age of $6\,\mathrm{Gyr}$, or desiccation of the atmosphere below a $3\,\mathrm{bar}$ surface pressure.

\subsection{Interior structure: \Zalmoxis}
\label{sec:methods:zalmoxis}

\Zalmoxis\footnote{\url{https://proteus-framework.org/Zalmoxis}} is a static one-dimensional radial structure solver for differentiated rocky planets and sub-Neptunes that returns the interior radius $R_\mathrm{int}$, the core-mantle boundary radius $R_\mathrm{cmb}$ and pressure $P_\mathrm{cmb}$, and the radial profiles of mass $M(r)$, density $\rho(r)$, gravity $g(r)$, pressure $P(r)$, and temperature $T(r)$.
Throughout the main text, $R_\mathrm{int}$ denotes the outer radius of the condensed (solid or molten) body at the surface pressure, as distinct from any photospheric or transit radius that includes the overlying atmosphere; the Results sections additionally normalise $R_\mathrm{int}$ to a molten reference state, defined separately for the standalone structure sweep ($R_0$, Figure~\ref{fig:shrinking}) and for the coupled run pairs ($R_\mathrm{base}$, Figure~\ref{fig:contraction_grid}).
The planet is assumed fully differentiated into an iron core and a silicate mantle from the outset, consistent with the rapid metal-silicate separation expected during accretion \citep{Kleine2009GeCoA735150K,Rubie2011EPSL30131R}; undifferentiated or partially mixed configurations are not treated.
Assuming hydrostatic and thermodynamic equilibrium, the solver integrates
\begin{equation}
\frac{dM}{dr} = 4 \pi r^2 \rho,
\quad
\frac{dg}{dr} = 4 \pi G \rho - \frac{2 g}{r},
\quad
\frac{dP}{dr} = -\rho \, g
\label{eq:zalmoxis_odes}
\end{equation}
from the centre to the surface, closed at each radial shell by a per-layer equation of state $\rho(P)$ or $\rho(P,T)$.
Layer membership is set by cumulative mass fractions.

For the coupled \PROTEUS\ runs reported in this work, \Zalmoxis\ uses the \PALEOS\ multiphase EoS framework of \citet{Attia2026subm} as its thermodynamic basis: the \texttt{PALEOS:iron} table for the core and the \texttt{PALEOS:MgSiO$_3$} table for the mantle.
The hydrostatic structure is solved against the unified \PALEOS\ silicate table, while the phase-specific property surfaces (density, heat capacity, thermal expansivity, and the adiabatic gradient) that the interior energetics module consumes are built from the separate \PALEOS\ solid and liquid tables.
This keeps the properties resolved across the melting-curve discontinuity that a single unified table interpolates through.
The silicate liquidus follows the Simon-Glatzel fit of \citet{Belonoshko2005PhRvL94s5701B} below 2.55\,GPa and the power-law fit of \citet{Fei2021NatCo12876F} above (a crossover in the shallow mantle, far below the $\sim$136\,GPa of Earth's core-mantle boundary), with the crossover pressure fixed by continuity.
The liquid branch follows the RTpress formulation of \citet{Wolf2018PEPI27859W}, whose larger thermal expansivity \citep{Stixrude2009EPSL278226S} is what lets a thermally hot, partially molten super-Earth have a different radius from its cold-isothermal counterpart at the same composition.
The solidus is taken as a fixed fraction of the liquidus, $T_\mathrm{sol}(P) = f\,T_\mathrm{liq}(P)$ with $f = 0.8$ an adopted constant fraction following the solidus-to-liquidus ratio of the \citet{Stixrude2014RSPTA37230076S} MgSiO$_3$ melting parametrisation, and the melt fraction follows from the lever rule between the two.
The mantle is treated as volatile-free in its structure and physical properties, which isolates the structural response from dissolved-volatile feedback for the comparisons reported here.
Dissolved volatiles are nonetheless a critical component of magma ocean structure and are probably central to the interior structure of volatile-rich sub-Neptunes \citep{Dorn2021ApJ922L4D,Boley2023ApJ954202B,McCormickKilbride2016NatCo713744K}; this feedback is not treated here and is developed in follow-up work (Attia et al., in preparation).

The initial thermal profile is anchored at the core-mantle boundary by setting
\begin{equation}
T_\mathrm{cmb} = T_\mathrm{liq}^\mathrm{PALEOS}\!\left(P_\mathrm{cmb}\right) + \Delta T_\mathrm{super},
\label{eq:zalmoxis_liquidus_super}
\end{equation}
where $T_\mathrm{liq}^\mathrm{PALEOS}$ is the \PALEOS\ silicate liquidus introduced above (the \citealt{Fei2021NatCo12876F} branch at core-mantle boundary pressures) and $\Delta T_\mathrm{super}$ is a configurable super-liquidus excess that sets the initial magma ocean thermal state ($500\,\mathrm{K}$ in the fiducial runs), and the adiabat is integrated upward from $(R_\mathrm{cmb}, T_\mathrm{cmb})$.
On the very first \PROTEUS\ iteration, before $P_\mathrm{cmb}$ is populated from a previous \Zalmoxis\ solution, it is estimated from the mass- and iron-fraction-aware scaling of \citet{Noack2020AA638A129N}; this avoids the systematic bias an Earth-only fixed value would introduce at super-Earth masses.

The radial domain is discretised on $N_r = 300$ grid points at super-Earth masses (and $150$ at $1\,\Mearth$), the ODEs of Equation~\ref{eq:zalmoxis_odes} are integrated with an adaptive Runge-Kutta scheme, and the outer mass-radius loop, a shooting method on the central pressure, is closed to a relative planet-mass mismatch below $10^{-3}$.
Inside the coupled loop the solver is not called at every step: a re-solve is triggered when the time since the last solve exceeds an update interval, when the relative change in magma-layer temperature, in the global melt fraction, or in the dissolved-volatile composition of the magma exceeds configured thresholds, or when a staleness ceiling is reached.
Each re-solve passes the updated structure to \Aragog\ and takes back the temperature profile \Aragog\ has evolved to that time.
This profile drives the structure solve in place of Equation~\ref{eq:zalmoxis_liquidus_super}, so at evolved times the thermal profile follows the cooling and partially crystallised mantle rather than the initial superheated anchor (Section~\ref{sec:methods:aragog_zalmoxis_coupling}).
A short pre-main-loop equilibration alternates between outgassing and \Zalmoxis\ until both $R_\mathrm{int}$ and $P_\mathrm{surf}$ have each converged to within $1\,\%$.

The configuration options, alternative EoS families, sensitivity-test settings, and the approximately 150-case automated verification suite (analytic limits, Preliminary Reference Earth Model (PREM) and mass-radius benchmarks, tolerance and grid convergence, numpy/JAX parity) are described in the \href{https://proteus-framework.org/Zalmoxis}{\Zalmoxis\ documentation}; the analytic and limiting-case tests are summarised in Appendix~\ref{sec:app:zalmoxis}, and the comparison against published super-Earth structure models up to $20\,\Mearth$ is reported in Appendix~\ref{sec:val:zalmoxis}.

\subsection{Interior energetics: \Aragog}
\label{sec:methods:aragog}

\Aragog\footnote{\url{https://proteus-framework.org/Aragog}} is a one-dimensional, spherically symmetric finite-volume solver for the radial thermal evolution of rocky planetary mantles in solid, fully molten, or partially molten states.
It is the entropy-based successor to the SPIDER mantle dynamics code \citep{Bower2018PEPI27449B,Bower2019AA631A103B,Bower2022PSJ393B}, written to reproduce the SPIDER magma ocean physics described below but evolved and expanded from that baseline.
The deviations from SPIDER, detailed in Sections~\ref{sec:methods:aragog:entropy} to~\ref{sec:methods:aragog:numerics}, include $C^\infty$ regularisations of the porosity and permeability factors that enter melt-solid separation, that is, replacements of these piecewise factors by infinitely differentiable (smooth) functions, which keep the solver Jacobian continuous across phase boundaries \citep{Spiegelman2016GGG172213S}, an analytic Jacobian (compiled with JAX), and integrator robustness fixes calibrated for super-Earth coupled runs.
The new Python implementation is the module used for all coupled \PROTEUS\ runs in this work.
Most of the governing equations already appear in \citet{Bower2018PEPI27449B}; we nonetheless set out the full formulation below because the subtle differences from it propagate critically into the interior-structure solution, and conveying the complete physical implementation is the central purpose of the reformulation.
The single prognostic variable is the specific entropy $S(r,t)$; temperature, density, melt fraction, heat capacity, thermal expansivity, and the adiabatic gradient are diagnostic and are recovered from the \PALEOS\ pressure-entropy tables of \citet{Attia2026subm} that are consistent with the per-shell EoS used by \Zalmoxis\ (Section~\ref{sec:methods:zalmoxis}).
The entropy formulation absorbs the latent heat of fusion into the entropy axis of the EoS table, so phase boundaries are crossed without the effective heat-capacity divergence that a temperature-based formulation must regularise across the solidus and the liquidus \citep{Bower2018PEPI27449B}.
\Aragog\ employs a mixing-length-theory (MLT) closure for convective heat transport \citep{Prandtl1925ZaMM5136P,Vitense1953ZA32135V,Abe1993Litho30223A}, in contrast to one-dimensional or zero-dimensional boundary-layer parameterisations that represent the convecting mantle by a single (or piecewise) temperature evolving under a Nusselt-Rayleigh scaling at an upper-boundary thermal layer \citep[e.g.][]{Solomatov2007eveabook91S,ElkinsTanton2008EPSL271181E,Lebrun2013JGRE1181155L,Schaefer2016ApJ82963S,Salvador2017JGRE1221458S,KrissansenTotton2022ApJ933115K}.
The MLT closure resolves the radial entropy profile, the propagating crystallisation front, the depth-dependent latent-heat release, and the interaction of conduction with convection inside the rheological transition; the price is a stiffer ODE system that must be integrated implicitly across the mush region, defined here as the radial interval where solid and liquid coexist ($0 < \phi < 1$, with $\phi$ the local melt mass fraction, between the local solidus and liquidus).

\subsubsection{Entropy balance and heat fluxes}
\label{sec:methods:aragog:entropy}

Each spherical shell, of constant mass by construction of the mass-coordinate mesh (Section~\ref{sec:methods:aragog:mesh}), between the core-mantle boundary at $r_\mathrm{cmb}$ and the surface at $r_\mathrm{top}$ obeys the integral entropy balance
\begin{equation}
\int_V \rho\,T \left.\frac{\partial S}{\partial t}\right|_\xi dV
= -\int_{\partial V} \mathbf{F}\cdot\mathbf{n}\, dS
+ \int_V \rho\,H\, dV,
\label{eq:aragog_entropy}
\end{equation}
where $\mathbf{F}\,[\mathrm{W\,m^{-2}}]$ is the radial heat flux, $H\,[\mathrm{W\,kg^{-1}}]$ the internal heating rate per unit mass, and the capacitance $\rho\,T$ multiplying $\partial S/\partial t$ replaces the temperature-form $\rho c_p$ and is consistent with the thermodynamic identity $T\,\mathrm{d}S = c_p\,\mathrm{d}T - (\alpha T/\rho)\,\mathrm{d}P$ enforced internally by the EoS lookups.
The time derivative is taken at constant mass coordinate $\xi$ (Section~\ref{sec:methods:aragog:mesh}); each finite-volume cell coincides with a material volume of constant mass and the net mass flux through cell faces vanishes.
The total radial heat flux is the sum of four contributions,
\begin{equation}
F_\mathrm{tot} = F_\mathrm{cond} + F_\mathrm{conv} + F_\mathrm{grav} + F_\mathrm{mix}.
\label{eq:aragog_Ftot}
\end{equation}

The conduction flux is rewritten in entropy-gradient form by combining the Fourier law with the thermodynamic identity $(\partial T/\partial S)_P = T/c_p$ and the EoS-tabulated isentropic temperature gradient $(\partial T/\partial P)_S$:
\begin{equation}
F_\mathrm{cond} = -k\,\left[\frac{T}{c_p}\,\frac{\partial S}{\partial r}
+ \left(\frac{\partial T}{\partial P}\right)_{\!S}\frac{\partial P}{\partial r}\right].
\label{eq:aragog_Fcond}
\end{equation}
When the entropy gradient vanishes the conduction flux reduces to its adiabatic part alone; the second term is therefore a residual conductive flux down the planetary adiabat.

The convective flux is parameterised as eddy diffusion of entropy,
\begin{equation}
F_\mathrm{conv} = \rho\,T\,\kappa_h\,\max\!\left(-\frac{\partial S}{\partial r},\,0\right),
\label{eq:aragog_Fconv}
\end{equation}
with the instability criterion $\partial S/\partial r < 0$.
Because the entropy gradient itself measures departures from the adiabat, no explicit superadiabatic-gradient subtraction is required; this structural simplification, inherited from \citet{Bower2018PEPI27449B}, distinguishes the entropy formulation from temperature-based MLT closures \citep[e.g.][]{Abe1993Litho30223A,Solomatov2007eveabook91S}.
The onset switch at $\partial S/\partial r = 0$ is implemented as a hard mask rather than as a smoothed sigmoid, because a sigmoid would leak convective transport into stably stratified cells; only the gradient magnitude used in the velocity scales is regularised at negligible amplitude so that the analytic Jacobian of Section~\ref{sec:methods:aragog:numerics} stays finite at exactly vanishing entropy gradients.
The eddy diffusivity $\kappa_h$ is set by the MLT closure of Section~\ref{sec:methods:aragog:eddy}.

In the partially molten regime, gravity drives a vertical separation of melt and solid, expressed as a melt mass flux
\begin{equation}
\begin{aligned}
j_\mathrm{grav} &= \rho\,\phi\,(1-\phi)\,v_\mathrm{rel}\,\mathrm{smth}(\phi), \\
v_\mathrm{rel} &= \frac{|\rho_m-\rho_s|\,g\,K(\zeta)}{\eta_m},
\end{aligned}
\label{eq:aragog_jgrav}
\end{equation}
where $\rho$ is the mixture density, $\rho_m$, $\rho_s$, and $\eta_m$ are the melt density, solid density, and melt dynamic viscosity, $K(\zeta)$ is the regime-dependent permeability defined in Section~\ref{sec:methods:aragog:eddy}, and $\mathrm{smth}(\phi)$ is the two-branch hyperbolic-tangent phase-boundary smoothing of \citet{Bower2018PEPI27449B},
\begin{equation}
\mathrm{smth}(\phi) = \begin{cases}
\tfrac{1}{2}\!\left[1 - \tanh\!\left(\dfrac{\phi - 1}{\Delta_\phi}\right)\right], & \phi > \tfrac{1}{2}, \\[6pt]
\tfrac{1}{2}\!\left[1 + \tanh\!\left(\dfrac{\phi}{\Delta_\phi}\right)\right], & \phi \le \tfrac{1}{2},
\end{cases}
\label{eq:aragog_smth}
\end{equation}
with width $\Delta_\phi = 0.01$ in the runs of this work, so that $\mathrm{smth}(\phi) \approx 1$ across the bulk of the mush region, decreasing to one half at the solidus and liquidus; the changeover at $\phi = \tfrac{1}{2}$ merely selects the nearer phase boundary for the smoothing and is not a rheological threshold.
The corresponding heat flux is
\begin{equation}
\begin{aligned}
F_\mathrm{grav} &= j_\mathrm{grav}\,L(P), \\
L(P) &= T_\mathrm{fus}(P)\,[S_\mathrm{liq}(P) - S_\mathrm{sol}(P)],
\end{aligned}
\label{eq:aragog_Fgrav}
\end{equation}
where $L(P)$ is the pressure-dependent latent heat of fusion read from the \PALEOS\ table.
The same $\mathrm{smth}$ factor, evaluated on the phase state of the cell below each interface, additionally suppresses $j_\mathrm{grav}$ where the lower neighbour is a pure phase, reflecting the physical constraint that melt cannot drain through a fully molten (or fully solid) lower mantle.

Compositional mixing of the melt fraction is implemented as a diffusive flux that relaxes the entropy gradient toward the local lever-rule prediction,
\begin{equation}
\begin{aligned}
F_\mathrm{mix} = -\kappa_c\,\rho\,T_\mathrm{fus}\,\mathrm{smth}(\phi)\Bigg[
&\frac{\partial S}{\partial r} - \overline{S'_P}\,\frac{\partial P}{\partial r}
\Bigg], \\
\overline{S'_P} \equiv\;
&\phi\,\frac{\partial S_\mathrm{liq}}{\partial P} + (1-\phi)\,\frac{\partial S_\mathrm{sol}}{\partial P},
\end{aligned}
\label{eq:aragog_Fmix}
\end{equation}
with $\kappa_c$ the compositional eddy diffusivity (Section~\ref{sec:methods:aragog:eddy}) and the bracketed expression the entropy-gradient excess relative to the linear lever-rule interpolation between the solidus and liquidus entropy gradients at the local pressure.
Inside the mush region the identity $\partial S/\partial r - [\phi\,\partial S_\mathrm{liq}/\partial P + (1-\phi)\,\partial S_\mathrm{sol}/\partial P]\,\partial P/\partial r = (S_\mathrm{liq} - S_\mathrm{sol})\,\partial \phi/\partial r$ holds, so $F_\mathrm{mix}/L(P)$ recovers the canonical melt mass flux $j_\mathrm{mix} = -\rho\,\kappa_c\,\partial \phi/\partial r$ of \citet{Abe1997PEPI10027A} and \citet{Solomatov2007eveabook91S}, modulo the mush-region smoothing factor (Equation~\ref{eq:aragog_smth}).

\subsubsection{Heat sources}
\label{sec:methods:aragog:heating}

Two contributions enter $H$:
\begin{equation}
H = H_\mathrm{radio} + H_\mathrm{tidal}.
\label{eq:aragog_H}
\end{equation}
The radiogenic heating is time-dependent and assumed spatially uniform across the mantle,
\begin{equation}
H_\mathrm{radio}(t) = \sum_i \chi_i\,\varphi_i\,\exp\!\left(-\frac{\ln 2\,(t - t_0)}{\tau^{1/2}_i}\right),
\label{eq:aragog_Hradio}
\end{equation}
with mass fraction $\chi_i$, specific power $\varphi_i$, and half-life $\tau^{1/2}_i$ for each isotope $i$; for the runs in this work the inventory comprises the long-lived isotopes $^{40}$K, $^{232}$Th, $^{235}$U, and $^{238}$U, with abundances prescribed at bulk-silicate-Earth concentrations at a reference age of $4.567\,\mathrm{Gyr}$ and decayed to the simulation epoch, following the \PROTEUS\ defaults of \citet{Lichtenberg2021JGRE12606711L,Nicholls2024JGRE12908576N}.
Tidal heating $H_\mathrm{tidal}$ is supported as an externally configured per-node profile, constant in time within an outer step but updated between outer steps from the self-consistent tidal-dissipation calculation in the \PROTEUS\ orbit module \citep[\texttt{lovepy};][]{Nicholls2025MNRAS5412566N,vanDijk2026PSJ794V}; it is set to zero in the runs of this work.
The $P\,\mathrm{d}V$ work performed when melt of different density from the local matrix is redistributed across the pressure gradient by gravitational separation or compositional mixing does not appear as a separate volumetric source.
Because the latent heat transported by $j_\mathrm{grav}$ and $j_\mathrm{mix}$, $L = T_\mathrm{fus}\,[S_\mathrm{liq} - S_\mathrm{sol}]$, equals the enthalpy difference between the coexisting phases along the melting curve, the divergence of the $L$-weighted mass fluxes in Equations~\ref{eq:aragog_Fgrav} and \ref{eq:aragog_Fmix} accounts for the volumetric work (dilatation heating) implicitly, consistent with the flux-form treatment of \citet{Bower2018PEPI27449B}.

\subsubsection{Mass coordinates}
\label{sec:methods:aragog:mesh}

The radial mesh is spaced uniformly in a mass coordinate $\xi$ defined by
\begin{equation}
\xi(r) = \left[3\int_{r_\mathrm{cmb}}^{r} \frac{\rho^*(r')}{\rho^*_\mathrm{mantle}}\,r'^2\,\mathrm{d}r' + \xi_\mathrm{cmb}^3\right]^{1/3},
\label{eq:aragog_xi}
\end{equation}
with $\rho^*(r)$ the configured pressure-density relation, $\rho^*_\mathrm{mantle}$ the volume-averaged density of the mantle shell, and $\xi_\mathrm{cmb} = r_\mathrm{cmb}$, so that $\xi_\mathrm{top} \equiv r_\mathrm{top}$.
Spatial gradients in mass-coordinate mode convert via the chain rule
\begin{equation}
\frac{\partial \psi}{\partial r} = \frac{\rho^*(r)}{\rho^*_\mathrm{mantle}}\left(\frac{r}{\xi}\right)^2\frac{\partial \psi}{\partial \xi}.
\label{eq:aragog_chainrule}
\end{equation}
The mass coordinate concentrates resolution where the density is high, which preserves resolution across the steep density jumps at the rheological transition and at the CMB that a uniform-radius mesh would underresolve.
The pressure-density relation is taken from the five-column mesh file written by \Zalmoxis\ ($r$, $P$, $\rho$, $g$, $T$; \Aragog\ consumes the first four); a configured Adams-Williamson profile can be used instead and underlies the SPIDER parity tests of Appendix~\ref{sec:val:aragog}.
Each \Zalmoxis\ structure update triggers a mesh rebuild that preserves the evolved entropy profile and is validated by an end-of-mantle radius check and a mantle mass-conservation check.

\subsubsection{Phase mixing and thermophysical properties}
\label{sec:methods:aragog:phase}

The melt mass fraction is recovered directly from the lever rule on the entropy axis of the \PALEOS\ table,
\begin{equation}
\phi(P,S) = \mathrm{clip}\!\left(\frac{S - S_\mathrm{sol}(P)}{S_\mathrm{liq}(P) - S_\mathrm{sol}(P)},\;0,\;1\right),
\label{eq:aragog_phi}
\end{equation}
with $S_\mathrm{sol}(P)$ and $S_\mathrm{liq}(P)$ tabulated alongside the phase EoS; the clip itself is hard, as in SPIDER.
The associated porosity $\zeta = (\rho_s - \rho)/(\rho_s - \rho_m)$ enters the gravitational-separation permeability law (Section~\ref{sec:methods:aragog:eddy}) and is instead computed with a $C^\infty$ square-root-based soft clip onto $[0,1]$ (width $10^{-3}$).
The soft clip is numerical in purpose: a hard clip makes the solver Jacobian discontinuous at each solidus or liquidus crossing and drops the integrator to first order, whereas the smooth form keeps the analytic Jacobian (Section~\ref{sec:methods:aragog:numerics}) continuous across phase transitions.

In the mush region, density follows the volume-additive harmonic-mean rule and the thermal conductivity $k_\mathrm{mix}$ the linear blend of the melt and solid values $k_m$ and $k_s$,
\begin{equation}
\frac{1}{\rho_\mathrm{mix}} = \frac{\phi}{\rho_m} + \frac{1-\phi}{\rho_s},
\qquad
k_\mathrm{mix} = \phi\,k_m + (1-\phi)\,k_s,
\label{eq:aragog_rhokmix}
\end{equation}
the heat capacity in the latent-heat blend follows from the tabulated phase entropies as
\begin{equation}
c_p^\mathrm{mix}(P,S) = \frac{S_\mathrm{liq}(P) - S_\mathrm{sol}(P)}{T_\mathrm{liq}(P) - T_\mathrm{sol}(P)}\,T_\mathrm{mid}(P),
\label{eq:aragog_cpmix}
\end{equation}
with $T_\mathrm{mid}$ the mid-mush temperature, and the thermal expansivity follows from $\alpha = \rho\,(\partial(1/\rho)/\partial T)|_P$ along the same isobaric path through the mush region.
The dynamic viscosity is blended in log-space across the rheological transition,
\begin{equation}
\begin{aligned}
\log_{10}\eta &= \log_{10}\eta_s + (\log_{10}\eta_m - \log_{10}\eta_s)\,w(\phi), \\
w(\phi) &= \tfrac{1}{2}\!\left[1 + \tanh\!\left(\frac{\phi - \phi_\mathrm{rheo}}{\Delta_\mathrm{rheo}}\right)\right],
\end{aligned}
\label{eq:aragog_visc}
\end{equation}
with $\phi_\mathrm{rheo}$ the critical melt fraction (default 0.5) and $\Delta_\mathrm{rheo}$ the transition width (default 0.2) consistent with \citet{Bower2018PEPI27449B,Solomatov2007eveabook91S}; laboratory and theoretical estimates place the rheological transition at melt fractions of about $0.4$ to $0.6$ \citep{Rosenberg2005JMetG2319R,Costa2009GGG103010C}.
A two-stage blend is applied to all thermophysical properties: a two-branch tanh mush-region weight $\mathrm{smth}(\phi)$ (Equation~\ref{eq:aragog_smth}, with the same form and width $\Delta_\phi = 0.01$ as the $j_\mathrm{grav}$ and $F_\mathrm{mix}$ smoother) first separates single-phase from mush-region contributions, after which the mush-region property is smoothly combined with the relevant single-phase value to ensure continuity across the solidus and the liquidus.

\subsubsection{Eddy diffusivity and permeability}
\label{sec:methods:aragog:eddy}

The MLT eddy diffusivity is the product of a mixing length $l(r)$ and a regime-dependent velocity scale.
Following \citet{Abe1993Litho30223A}, \Aragog\ blends the viscous and inviscid limits via a $\tanh$ switch in the cell Reynolds number,
\begin{equation}
\begin{aligned}
v_\mathrm{visc} &= \frac{\alpha\,g\,T\,(-\partial S/\partial r)\,l^3}{18\,\nu\,c_p}, \\
v_\mathrm{inv} &= \frac{l}{4}\!\left[\frac{\alpha\,g\,T\,(-\partial S/\partial r)}{c_p}\right]^{1/2},
\end{aligned}
\label{eq:aragog_vels}
\end{equation}
\begin{equation}
\begin{aligned}
\kappa_h &= l\,[(1-w(\mathrm{Re}))\,v_\mathrm{visc} + w(\mathrm{Re})\,v_\mathrm{inv}], \\
w(\mathrm{Re}) &= \tfrac{1}{2}\!\left[1 + \tanh\!\left(\frac{\mathrm{Re} - \mathrm{Re}_\mathrm{crit}}{\Delta_\mathrm{Re}}\right)\right],
\end{aligned}
\label{eq:aragog_kappah}
\end{equation}
with $\mathrm{Re} = v_\mathrm{visc}\,l/\nu$, $\nu = \eta/\rho$ the kinematic viscosity, $\mathrm{Re}_\mathrm{crit} = 9/8$ (the critical value of \citealt{Abe1993Litho30223A}), and a narrow blend $\Delta_\mathrm{Re} = 0.01\,\mathrm{Re}_\mathrm{crit}$ that confines the inviscid scaling to the convecting regime; widening it leaks inviscid diffusion into the solid phase and induces $T_\mathrm{cmb}$ bistability \citep{Bower2018PEPI27449B}.
The mixing length is the distance to the nearest boundary, $l(r) = \min(r_\mathrm{top}-r,\;r-r_\mathrm{cmb})$.
A phase-modulated floor $\kappa_h \to \max(\kappa_h,\,\kappa_h^\mathrm{floor}\,w(\phi))$, with $w(\phi)$ the $\tanh$ weight of Equation~\ref{eq:aragog_visc}, activates in partially and fully molten regions ($\phi \gtrsim \phi_\mathrm{rheo}$), where a near-adiabatic entropy profile can otherwise drive the MLT diffusivity toward zero and stall the integration, and vanishes in solid regions so that no spurious convective flux is introduced; the adopted floor is $\kappa_h^\mathrm{floor} = 10\,\mathrm{m^2\,s^{-1}}$.
The compositional eddy diffusivity $\kappa_c$ in Equation~\ref{eq:aragog_Fmix} is taken proportional to the unfloored MLT diffusivity with a configurable scale factor (unity in the runs of this work).

The permeability factor $K$ in Equation~\ref{eq:aragog_jgrav} depends on the porosity and on the flow regime.
The three branches considered individually are
\begin{equation}
K(\zeta) = \begin{cases}
\tfrac{2}{9}\,a^2, & \zeta > 0.7715, \\[2pt]
\tfrac{5}{7}\,a^2\,\zeta^{4.5}, & 0.0769 \le \zeta \le 0.7715, \\[2pt]
10^{-3}\,a^2\,\dfrac{\zeta^2}{(1-\zeta)^2}, & \zeta < 0.0769,
\end{cases}
\label{eq:aragog_K}
\end{equation}
following \citet{Abe1993Litho30223A}, with $a$ the configured grain size.
The three regimes correspond to a melt-rich Stokes regime in which solid grains settle through liquid, a mid-porosity Rumpf-Gupte regime in which residual melt drains through a partially connected pore network, and a solid-rich Blake-Kozeny-Carman regime in which the remaining melt percolates through narrow throats of a granular solid framework.
\Aragog\ does not switch between branches at the regime boundaries but blends them with a $\tanh$ switch (widths 0.02 and 0.05 in $\zeta$) at the analytical-equality porosities, the Blake-Kozeny-Carman to Rumpf-Gupte boundary at $\zeta_1 = 0.0769$ and the Rumpf-Gupte to Stokes boundary at $\zeta_2 = 0.7715$, which keeps $K(\zeta)$ and its derivative continuous and is what the solver Jacobian sees; this is a regularisation update over the piecewise SPIDER form.

\subsubsection{Boundary conditions and core cooling}
\label{sec:methods:aragog:bc}

The surface boundary condition is taken in flux-prescribed mode in the coupled \PROTEUS\ runs of this work: at every outer step the outgoing radiative flux $F_\mathrm{atm}$ from \AGNI\ is imposed at the topmost basic node (cell face, the uppermost layer edge), ensuring a closed mantle-atmosphere energy balance and removing surface-temperature drift between modules \citep{Nicholls2024JGRE12908576N}.
This boundary condition imposes the instantaneous atmospheric flux on the interior at every step and does not require the interior heat flow $F_\mathrm{int}$ to match it; the balance $F_\mathrm{atm} \simeq F_\mathrm{int}$ is reached only at the quasi-steady endpoint of Section~\ref{sec:results:volatiles}.
No parameterised conductive boundary layer is imposed at the surface: the prescribed atmospheric flux closes the energy balance at the top face directly, and any conductive lid that develops as the near-surface material solidifies is represented by the resolved entropy profile at mesh resolution rather than by a sub-grid boundary-layer model.
At the core-mantle boundary we use the energy-balance closure of \citet{Bower2018PEPI27449B}, in which the core enthalpy budget
\begin{equation}
(\rho c_p V)_\mathrm{core}\,\frac{\mathrm{d}T_\mathrm{core}}{\mathrm{d}t} = -A_\mathrm{cmb}\,F_\mathrm{cmb}
\label{eq:aragog_corebalance}
\end{equation}
is closed against the lowermost mantle cell through the linear scaling $T_\mathrm{core} \simeq \hat{T}_\mathrm{core}\,T_1$ ($\hat{T}_\mathrm{core} = 1.147$, after \citealt{Bower2018PEPI27449B}).
The boundary entropy gradient $\partial S/\partial r|_\mathrm{cmb}$ is appended as an additional state variable and evolved consistently with this core enthalpy budget at each ODE step, which avoids the numerical stiffness that a fixed-flux closure would impose on the lowermost cell during early-stage core cooling \citep{Bower2018PEPI27449B}.

\subsubsection{Numerical method}
\label{sec:methods:aragog:numerics}

Equation~\ref{eq:aragog_entropy} is discretised by finite volumes with the fluxes at $N = 80$ basic cell-face nodes (default) and the entropy at the $N-1 = 79$ staggered cell-centred nodes.
The semi-discrete form, for staggered node $i$, is
\begin{equation}
(\rho\,T\,V)_i\,\frac{\mathrm{d}S_i}{\mathrm{d}t}
= -F_{i+1/2}\,A_{i+1/2} + F_{i-1/2}\,A_{i-1/2} + \rho_i\,H_i\,V_i,
\label{eq:aragog_fvm}
\end{equation}
with $A_{i\pm 1/2} = 4\pi r^2_{i\pm 1/2}$ and $V_i = \tfrac{4}{3}\pi(r^3_{i+1/2} - r^3_{i-1/2})$.
Basic-node values follow from linear interpolation between the bracketing staggered nodes and entropy gradients from centred differences in uniform $\xi$-space, chain-ruled through Equation~\ref{eq:aragog_chainrule}.

The resulting stiff ODE system is integrated with an implicit, variable-order stiff solver \citep{Hindmarsh} and an analytic Jacobian compiled with JAX \citep{Bradbury2018jax}.
The $C^\infty$ porosity and permeability regularisations of Sections~\ref{sec:methods:aragog:phase} and~\ref{sec:methods:aragog:eddy}, together with the tanh property blends, keep the flux differentiable through the phase transitions; where a step nonetheless fails to converge within the mush region, the coupling wrapper retries it with a reduced timestep and relaxed tolerance so that crystallisation is traversed cleanly.
\Aragog\ then returns the updated radial profiles and per-component fluxes to the next outer coupling step.

\subsubsection{Verification}
\label{sec:methods:aragog:verification}

The full physical and numerical formulation is documented on the model description, energy-equation, heat-transport, and phase-transition pages of the \Aragog\ documentation.
\Aragog\ includes an automated verification suite covering EoS lookups, phase-evaluator parity, conservation-law checks (energy, mass, grey-body radiative balance), JAX-vs-numpy parity, and mesh-gravity fallback consistency.
First-principles validation against published magma ocean evolution benchmarks and parity comparisons with \texttt{SPIDER} for the anchor compositions of Section~\ref{sec:methods:anchors} are reported in Appendix~\ref{sec:app:aragog} and Appendix~\ref{sec:val:aragog}.

\subsection{Coupling of \Aragog\ and \Zalmoxis\ across the mush region}
\label{sec:methods:aragog_zalmoxis_coupling}

At every outer \PROTEUS\ step \Zalmoxis\ supplies the static structural basis ($r$, $P$, $\rho$, $g$, $T$ on the five-column mesh of Section~\ref{sec:methods:zalmoxis}) that sets \Aragog's mass coordinate, gravity profile, and the pressure axis of the \PALEOS\ tables; \Aragog\ in turn returns the entropy and melt-fraction profiles and the evolved temperature profile that the next \Zalmoxis\ solve uses in place of Equation~\ref{eq:zalmoxis_liquidus_super} and for the per-shell solid/liquid \PALEOS\ branch selection, following the structure-update criteria of Section~\ref{sec:methods:zalmoxis}.

The mush region ($0 < \phi(r,t) < 1$) is the natural meeting point of the two modules: \Aragog\ derives $\phi$ from the lever rule (Equation~\ref{eq:aragog_phi}) at the local $(P, S)$, with $S$ evolved by Equation~\ref{eq:aragog_entropy} and $P(r)$ supplied by \Zalmoxis, so the position and width of the mush region depend jointly on both fields and a structural update at fixed $S(r)$ can shift it radially by moving the $P$-axis under the solidus and liquidus.

The feedback runs both ways: inside the mush region the density follows the harmonic-mean rule of Equation~\ref{eq:aragog_rhokmix} between the solid \PALEOS\ branch and the more thermally expansive liquid branch, so a partially molten mantle is systematically less dense than its fully-solid counterpart at the same composition and pressure, and at the next \Zalmoxis\ trigger this $\rho_\mathrm{mix}(r)$ resets $R_\mathrm{int}$, $g(r)$, $P(r)$, and $P_\mathrm{cmb}$, which in turn shift the solidus and liquidus and the mush region itself; the self-consistent magma ocean radius is the converged solution of this loop, not an a-priori anchor.

\subsection{Outgassing: \atmodeller\ and the \CALLIOPE\ cross-validation reference}
\label{sec:methods:atmodeller}

Volatile partitioning between the atmosphere and the magma ocean at the surface boundary is solved at thermochemical-solubility equilibrium at every outer \PROTEUS\ step.
The primary outgassing module is \atmodeller\ \citep{Bower2025ApJ99559B}, with the \CALLIOPE\ module \citep{Bower2022PSJ393B,Nicholls2024JGRE12908576N,Nicholls2025MNRAS5412566N,Nicholls2026NatAs10809N,Sossi2023EPSL60117894S} retained as a cross-validation reference.
Both modules take as input the current elemental mass budgets of H, C, N, S, which evolve in time as escape removes volatiles (Section~\ref{sec:methods:proteus}), and the surface temperature delivered by \Aragog, and return the surface partial pressures of all gas species (the species sets are listed below) and the dissolved-in-melt mass of each volatile.
The oxidation state of the equilibrium is set in one of two modes, which are dual formulations of the same chemistry and differ in which quantity is prescribed and which is solved for; the fixed-fugacity mode reproduces the treatment standard in earlier coupled models, while the oxygen-conserving mode implements the conserved-oxygen accounting motivated in Section~\ref{sec:intro}, and comparing the two quantifies the consequences of that modelling choice.
Iron is the most abundant element in rocky planets with more than one oxidation state, so the equilibrium between its metal and oxide phases regulates the oxygen fugacity of the silicate melt, and offsets from the iron-w\"ustite (IW) equilibrium are the natural reference scale for mantle oxidation \citep{ONeill1993CoMP114296O,Hirschmann2021GeCoA31374H}.
In the fixed-fugacity mode the oxygen fugacity is buffered to the IW equilibrium with a configurable IW-relative shift, the treatment employed in all previous \PROTEUS\ applications \citep{Nicholls2024JGRE12908576N,Nicholls2025MNRAS5362957N,Nicholls2025MNRAS5412566N,Nicholls2026NatAs10809N,Calder2026MNRAS549g1007C,Lichtenberg2026PSJ7108L,Sastre2026arXiv260620249S,vanDijk2026PSJ794V,Postolec2026arXiv260715011P}; \CALLIOPE\ implements the IW parameterisation of \citet{Fischer2011EPSL304496F}, whereas \atmodeller\ uses the composite calibration of \citet{ONeill1993CoMP114296O} and \citet{Hirschmann2021GeCoA31374H}; the two parameterisations agree in $\log_{10}\fO$ to within $\sim$0.2\,dex across magma ocean temperatures (Figure~\ref{fig:val_calliope_atmodeller}), so the cross-module comparison at a fixed IW shift probes the formulation difference rather than the buffer choice.
The equilibrium in this mode is the root of the elemental mass-balance system
\begin{equation}
r_e = m_e^\mathrm{atm} + m_e^\mathrm{melt} - m_e^\mathrm{target} = 0,
\qquad e \in \{\mathrm{H, C, N, S}\},
\label{eq:outgas_massbal}
\end{equation}
where $m_e^\mathrm{atm}$ and $m_e^\mathrm{melt}$ are the atmospheric column mass and the dissolved mass of element $e$, both functions of the surface partial pressures and of the imposed oxygen fugacity
\begin{equation}
\log_{10} \fO = \log_{10} \fO^\mathrm{IW}(T) + \Delta\mathrm{IW},
\label{eq:outgas_fo2}
\end{equation}
which fixes the O$_2$ fugacity itself; together with the temperature-dependent equilibrium constants this sets the abundance ratio of the reduced and oxidised member of each redox pair (for example H$_2$ to H$_2$O and CO to CO$_2$) and enters the $\fO$-dependent solubility laws.
The planetary oxygen mass is then a determined output rather than an input: at a given surface temperature and H, C, N, S budget the buffered chemistry fixes the total volatile-bound oxygen.

In the oxygen-conserving mode, introduced in this work and used for the oxygen-conserving arm of the redox comparison (Section~\ref{sec:results:redox_mode}), the oxygen budget is instead prescribed as a fifth conserved elemental inventory: the residual vector of Equation~\ref{eq:outgas_massbal} is extended to $e \in \{\mathrm{H, C, N, S, O}\}$ and $\Delta\mathrm{IW}$ becomes a solved-for quantity.
The conserved quantity is the free oxygen, the oxygen held in the atmosphere and dissolved in the melt as volatile species, and not the oxygen locked in the silicate oxides.
For each oxygen-conserving run this budget is taken from the fixed-fugacity run of the same inputs, its twin, at the initial outgassing equilibrium, and then held fixed as the planet evolves.
In \CALLIOPE\ the solution vector becomes
\begin{equation}
\mathbf{x} = \left(p_\mathrm{H_2O},\; p_\mathrm{CO_2},\; p_\mathrm{N_2},\; p_\mathrm{S_2},\; \Delta\mathrm{IW}\right),
\label{eq:outgas_authO}
\end{equation}
while \atmodeller\ equivalently replaces its O$_2$ fugacity constraint by an oxygen mass constraint in the extended law-of-mass-action (xLMA) system; in both modules no fugacity condition is imposed and the surface oxidation state becomes a derived diagnostic of the solution rather than an input.
The closure is well posed because the total (atmospheric plus dissolved) oxygen mass increases monotonically with $\Delta\mathrm{IW}$ across the physically relevant redox range: a more oxidising state binds more H into H$_2$O at fixed hydrogen budget and more C into CO$_2$ at fixed carbon budget, and shifts nitrogen from the melt into atmospheric N$_2$ \citep{Dasgupta2022GeCoA336291D}, so each feasible oxygen target corresponds to exactly one fugacity offset.
This one-to-one mapping holds at a fixed temperature and volatile budget.
A given oxygen inventory does not fix the surface fugacity on its own, so as the mantle crystallises the derived $\fO$ of an oxygen-conserving run can diverge from the buffered value of its twin, most strongly at the oxidising end (Section~\ref{sec:results:redox_mode}).
The two modes invert one another, and their equivalence is demonstrated across the full $\Delta\mathrm{IW}$ sweep in Section~\ref{sec:results:redox_mode}.

\atmodeller\ implements the xLMA framework.
The reaction set is not hard-coded but constructed by Gaussian elimination of the species formula matrix at solver initialisation, so the solver scales transparently to arbitrary species sets, with equilibrium-constant data drawn from JANAF tables as compiled in \citet{Bower2025ApJ99559B}.
In the grid runs of this work \atmodeller\ includes every C-H-O-N-S gas species whose elemental budget exceeds a mass threshold; for the fiducial inventory this is H$_2$O, H$_2$, CO$_2$, CO, CH$_4$, N$_2$, S$_2$, SO$_2$, H$_2$S, and NH$_3$, together with free O$_2$, while rock-vapour species remain inactive, and graphite (C$_{(\mathrm{cr})}$) is the default condensate.
Solubility laws are species-specific: H$_2$O follows the peridotite calibration of \citet{Sossi2023EPSL60117894S}; CO$_2$ follows \citet{DIXON1995JPet361607D}; H$_2$ follows \citet{Hirschmann2012EPSL34148H}; N$_2$ follows \citet{Dasgupta2022GeCoA336291D}; S$_2$ follows \citet{Boulliung2023CoMP17856B}; CO follows \citet{Yoshioka2019GeCoA259129Y}; CH$_4$ follows \citet{Ardia2013GeCoA11452A}.
The package also offers a real-gas equation-of-state library (compensated Redlich-Kwong, virial, Beattie-Bridgeman, and ab-initio forms) for non-ideal corrections.
The simulation grid retains the ideal-gas closure, which is adequate across most of the parameter space but becomes an approximation at the highest-pressure corners, the oxidising end of the redox sweep and the most massive volatile-rich planets (Section~\ref{sec:results:volatiles}), where the surface pressure comes within a factor of a few of the $10^{5}\,\mathrm{bar}$ validity limit.

\CALLIOPE, the cross-validation reference inherited from \citet{Bower2022PSJ393B}, \citet{Sossi2023EPSL60117894S}, \citet{Nicholls2024JGRE12908576N}, \citet{Nicholls2025MNRAS5412566N}, and \citet{Nicholls2026NatAs10809N}, shares with \atmodeller\ both redox modes and the H$_2$O, CO$_2$, CH$_4$, and N$_2$ solubility laws, but takes its sulfur (S$_2$) solubility from \citet{Gaillard2022EPSL57717255G} and its CO solubility from \citet{Armstrong2015GeCoA171283A}, in place of the \citet{Boulliung2023CoMP17856B} and \citet{Yoshioka2019GeCoA259129Y} laws that \atmodeller\ uses.
Beyond the solubility laws, \CALLIOPE\ differs structurally in three ways.
First, the equilibrium system is reduced compared with that used by \atmodeller: four primary oxidised species (H$_2$O, CO$_2$, N$_2$, S$_2$) are the independent unknowns and the reduced species H$_2$ and CO are derived analytically from JANAF equilibrium constants at the surface $T$ and $\fO$, and CH$_4$ from a separate thermochemical fit \citep[][Eqs.~4--6]{Bower2022PSJ393B}.
Second, \CALLIOPE\ assumes ideal gas throughout and does not solve for condensation; separately, its dissolved H$_2$ budget is set to zero rather than to a solubility law, a mild simplification under the conditions considered in this work, since molecular hydrogen dissolves far more weakly than water \citep{Hirschmann2012EPSL34148H}.
Third, \CALLIOPE\ solves the equilibrium with a multi-start root-finder, whereas \atmodeller\ solves the same system deterministically.
For the cross-validation the two modules are run on a common eight-species set (H$_2$O, CO$_2$, N$_2$, S$_2$, H$_2$, CH$_4$, CO, SO$_2$) so that the comparison isolates the formulation difference rather than the species composition.

\subsection{Anchor compositions and parameter sweeps}
\label{sec:methods:anchors}
The simulation grid probes how the coupled structural and thermal evolution of a rocky super-Earth depends on its mass, its volatile inventory, and its oxidation state, organised as one-parameter sweeps around a single fiducial reference state.
The reference planet is a $5\,\Mearth$ body with an Earth-like core-mass fraction of $0.325$ \citep{Zeng2016ApJ819127Z}, hosted by a Sun-like star ($1\,M_\odot$) at $1\,\mathrm{AU}$, at a fiducial oxygen fugacity of $\Delta\mathrm{IW}=+4$, similar to Earth's present-day upper mantle \citep{Frost2008AREPS36389F,Sossi2020SciA61387S}, and started fully molten from the super-liquidus initial profile.
Throughout, the interior structure is computed for a volatile-free mantle: the dissolved volatile mass is subtracted from the atmospheric budget but does not enter the silicate density, so the structure responds to temperature and melt fraction alone.
This is a significant limitation: dissolved volatiles can substantially alter the radius and interior structure of magma ocean planets \citep{Dorn2021ApJ922L4D}.
We adopt the volatile-free treatment deliberately, however, to isolate the structural response to the rock-mantle phase changes that are the subject of this work; the dissolved-volatile feedback is the focus of a dedicated follow-up study (Attia et al., in preparation).

The fiducial volatile inventory at $1\,\Mearth$ is a deliberately volatile-poor endowment: three terrestrial oceans of hydrogen ($4.7\times10^{20}\,\mathrm{kg}$) and $2.73\times10^{20}\,\mathrm{kg}$ carbon, close to terrestrial bulk-silicate values \citep{Marty2012EPSL31356M,Hirschmann2018EPSL502262H}, together with $5\times10^{18}\,\mathrm{kg}$ nitrogen and $5\times10^{19}\,\mathrm{kg}$ sulfur, both set below the bulk-silicate-Earth estimates \citep{Krijt2023ASPC5341031K}, the sulfur by about a factor of twenty.
We adopt four prescriptions for how this inventory scales with planet mass (Table~\ref{tab:volatile_treatments}), spanning the range of volatile endowments that planet formation can imprint on rocky planets \citep{Lichtenberg2019NatAs3307L,Lichtenberg2021ApJ913L20L,Lichtenberg2022ApJ938L3L,Drazkowska2023ASPC534717D,Krijt2025ApJ990L72K}: a fixed absolute budget, for which the volatile mass fraction falls as $M^{-1}$; a mass-scaled budget that holds the volatile mass fraction constant; a hydrogen-enriched budget in which the hydrogen mass grows as $M^{3}$, so that its mass fraction rises as $M^{2}$ to about $1\,\%$ of the planet mass at $10\,\Mearth$, a proxy for the retained nebular hydrogen that more massive cores can hold \citep{Lee2015ApJ81141L,Ginzburg2016ApJ82529G,Owen2017ApJ84729O,Calder2026MNRAS549g1007C}, though helium and the self-gravity of the envelope are not treated; and a volatile-rich budget in which the entire inventory grows as $M^{2.5}$, representing volatile-rich and water-world formation \citep{Zeng2019PNAS1169723Z,Kimura2020MNRAS4963755K,Mousis2020ApJ896L22M,Kite2021ApJ909L22K,Luque2022Sci3771211L,Kimura2026ApJ1000220K}.
The two enriched prescriptions are bounded by the requirement that the surface pressure remain below $10^{5}\,\mathrm{bar}$, the validity limit of the outgassing thermodynamics; both reach about $7\times10^{4}\,\mathrm{bar}$ at $10\,\Mearth$ at the reducing end of the oxygen-fugacity range, and steeper scalings are excluded.

\begin{deluxetable}{lll}
\tablecaption{Volatile-inventory treatments and their scaling with planet mass.\label{tab:volatile_treatments}}
\tabletypesize{\footnotesize}
\tablehead{\colhead{Volatile treatment} & \colhead{Scaling} & \colhead{State at $10\,\Mearth$}}
\startdata
Fixed absolute & constant mass & fraction $\propto M^{-1}$ \\
Mass-scaled & all $\propto M$ & constant fraction \\
Hydrogen-enriched & H\,$\propto M^{3}$, rest $\propto M$ & $\sim 1\,\%$ hydrogen \\
Volatile-rich & all $\propto M^{2.5}$ & $\sim 0.4\,\%$ volatiles \\
\enddata
\end{deluxetable}

From the reference state we vary one axis at a time.
The planet mass spans $1$, $3$, $5$, and $10\,\Mearth$, while the interior-structure solver is validated against published models to $20\,\Mearth$ in Appendix~\ref{sec:val:zalmoxis}.
The mantle oxygen fugacity is sampled in two modes: the fixed-fugacity mode, in which the iron-w\"ustite offset is imposed, is run at thirteen levels from $\Delta\mathrm{IW}=-6$ to $+6$; the oxygen-conserving mode (Section~\ref{sec:methods:atmodeller}), in which oxygen is conserved alongside hydrogen, carbon, nitrogen, and sulfur and the offset becomes a derived diagnostic, is run at the twelve paired levels from $-6$ to $+5$.
Each oxygen-conserving run conserves the free oxygen budget calculated from its fixed-fugacity twin, so the quoted $\Delta\mathrm{IW}$ of an oxygen-conserving run labels the twin pairing rather than an imposed quantity, and the two modes are compared at equivalent oxygen budgets.
The $+6$ level has no oxygen-conserving counterpart because the fixed-fugacity equilibrium there demands an atmosphere more massive than the planet and so defines no finite oxygen budget to conserve (Section~\ref{sec:results:redox_mode}).
The interior structure is either recomputed as the mantle crystallises, the default in this work, or held fixed at the hot molten start, which isolates the contribution of crystallisation-driven contraction to the radius evolution.
The host star is taken to be the Sun or a generic $0.2\,M_\odot$ M dwarf with a PHOENIX spectrum \citep{Husser2013AA553A6H}, and the instellation is varied with orbital distance ($0.1$ to $1$\,AU) into the close-in, strongly irradiated regime.

Each simulation in the sweep is run with the full module stack: \Zalmoxis\ (interior structure), \Aragog\ (interior energetics), \AGNI\ (atmosphere), \texttt{MORS} (stellar evolution), \texttt{ZEPHYRUS} (atmospheric escape), and \atmodeller\ (outgassing), integrated until the mantle solidifies ($\Phi < 0.01$), a radiative quasi-steady state is reached, the age reaches $6\,\mathrm{Gyr}$, or the atmosphere is desiccated below a $3\,\mathrm{bar}$ surface pressure, whichever occurs first.
The mass and structure-update axes underpin the radius-contraction result of Section~\ref{sec:results:baseline}, the oxygen-fugacity and redox-mode axes the redox study of Section~\ref{sec:results:redox_mode}, and the volatile-inventory axis the composition study of Section~\ref{sec:results:volatiles}; the host-star, irradiation, and core-mass-fraction variations are collected as robustness checks (Section~\ref{sec:results:stellar}).

\section{Framework validation}
\label{sec:framework_validation}

Establishing the coupled framework is the first aim of this work (Section~\ref{sec:intro}).
Its interior-structure, mantle-energetics, and outgassing modules are new or substantially rebuilt, so before applying them across the super-Earth regime we verify each against independent references and the fully coupled framework against a community intercomparison.
We present here the three validations that most directly underpin the results: the interior-structure solver against published mass-radius models, the two outgassing modules against each other, and the fully coupled framework against the community magma ocean intercomparison.
The complete module-level test suite, the analytic and limiting-case tests, the parity of the mantle solver against SPIDER, and the interior-profile and speciation comparisons, is collected in Appendix~\ref{sec:validation}.

\subsection{Interior structure against published models}
\label{sec:val:main:structure}

\begin{figure}
\centering
\includegraphics[width=\columnwidth]{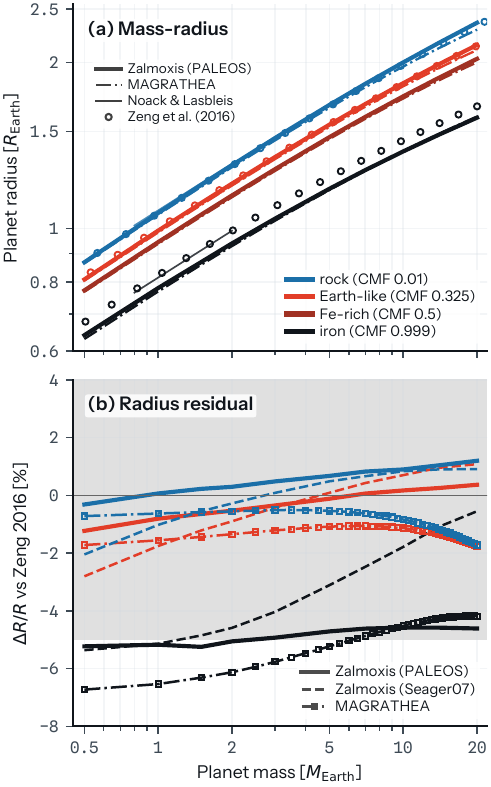}
\caption{
Interior-only mass-radius validation of the \Zalmoxis\ solver against published rocky-planet models, coloured by core-mass fraction (CMF).
\textbf{(a)} Mass-radius relations: \Zalmoxis\ with the PALEOS tables (solid), MAGRATHEA (dash-dot), the \citet{Noack2020AA638A129N} scaling (thin; shown over its $0.8$ to $2\,\Mearth$ calibration range), and the \citet{Zeng2016ApJ819127Z} tables (circles).
\textbf{(b)} Fractional radius residual against the \citet{Zeng2016ApJ819127Z} baseline for \Zalmoxis\ (PALEOS), \Zalmoxis\ with the \citet{Seager2007ApJ6691279S} equation of state, and MAGRATHEA, with the $\pm 5\,\%$ band shaded.
The Fe-rich (CMF $0.50$) case has no \citet{Zeng2016ApJ819127Z} tabulated baseline and so appears only in panel (a).
}
\label{fig:val_zalmoxis_mr}
\end{figure}

The interior-structure solver reproduces published rocky mass-radius relations across composition and mass (Figure~\ref{fig:val_zalmoxis_mr}).
Computing bare, condensed mass-radius relations from $0.5$ to $20\,\Mearth$ at four core-mass fractions, and comparing against the \citet{Zeng2016ApJ819127Z} tables, the independent MAGRATHEA code \citep{Huang2022MNRAS5135256H}, the \citet{Seager2007ApJ6691279S} equation of state, and the \citet{Noack2020AA638A129N} scaling relations, the \Zalmoxis\ radii agree with \citet{Zeng2016ApJ819127Z} to better than $1.3\,\%$ for rock and Earth-like compositions and to within about $5\,\%$ for the pure-iron endmember, where the reference equations of state themselves disagree most (Appendix~\ref{sec:val:zalmoxis} gives the composition-by-composition comparison and the interior profiles).
The agreement holds across the full super-Earth mass range, above the $\sim 2\,\Mearth$ calibration limit of the analytic scaling relations, which is the regime in which the coupled evolution requires the full structure solver.

\subsection{Outgassing modules against each other}
\label{sec:val:main:outgassing}

\begin{figure}
\centering
\includegraphics[width=\columnwidth]{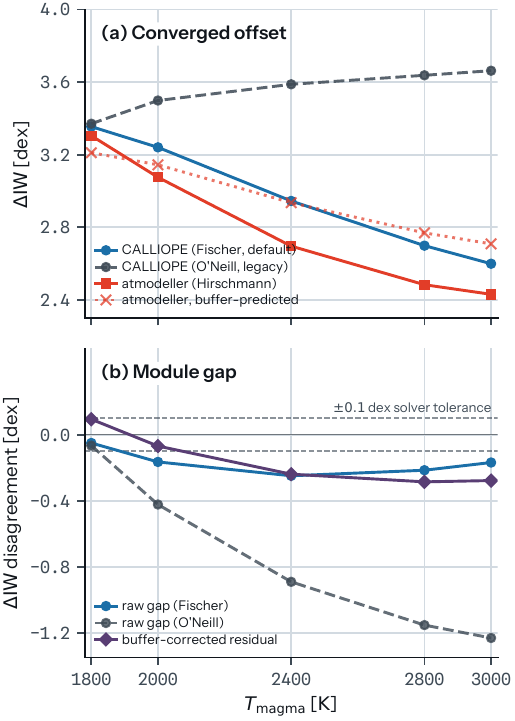}
\caption{
Agreement between the two \PROTEUS\ outgassing modules, \CALLIOPE\ and \atmodeller, through the shared oxygen-conserving entry point at the Earth bulk-silicate volatile inventory \citep{Krijt2023ASPC5341031K} with the volatile-oxygen reference set at $\Delta\mathrm{IW} = +3.5$ and melt fraction unity.
\textbf{(a)} Converged oxygen-fugacity offset $\Delta\mathrm{IW}$ against magma temperature for \CALLIOPE\ with the \citet{Fischer2011EPSL304496F} buffer (default) and the legacy \citet{ONeill2002ChGeo186151O} buffer, \atmodeller\ with its Hirschmann composite \citep{ONeill1993CoMP114296O,Hirschmann2021GeCoA31374H}, and the offset \atmodeller\ would show if the buffer were the only difference (dotted).
\textbf{(b)} Raw cross-module differences in oxygen fugacity for both \CALLIOPE\ buffers and the residual after the analytic buffer offset is removed; dashed lines mark the $\pm0.1\,\mathrm{dex}$ solver tolerance.
The Fischer-default difference stays within $0.25\,\mathrm{dex}$ across the range; the buffer-corrected residual measures the difference in the chemistry (solubility laws and equilibrium constants) once the buffer offset is removed, and is set by the sulfur solubility law.
}
\label{fig:val_calliope_atmodeller}
\end{figure}

The outgassing treatment is validated by cross-comparing the two independent \PROTEUS\ modules, \CALLIOPE\ and \atmodeller\ (Figure~\ref{fig:val_calliope_atmodeller}).
Inverting the same oxygen-conserving closure at the Earth bulk-silicate inventory \citep{Krijt2023ASPC5341031K}, the two solvers, which differ in their oxygen-fugacity buffer, solubility laws, equilibrium-constant fits, and solver architecture, and are both run here in their common ideal-gas closure, agree in the derived oxygen fugacity to within $0.25\,\mathrm{dex}$ across the magma ocean temperature range.
The small residual is set by the buffer convention at lower temperature and by the sulfur solubility law at the hottest, most oxidising end, where the sulfate-regime difference between the \citet{Gaillard2022EPSL57717255G} and \citet{Boulliung2023CoMP17856B} laws is largest (Appendix~\ref{sec:app:calliope_atmodeller}), and both produce a mantle $\fO$ consistent with Earth's modern upper mantle \citep{Frost2008AREPS36389F}, so the cross-module difference is small compared with the uncertainty on estimates of Earth's upper-mantle $\fO$.
The dotted curve in Figure~\ref{fig:val_calliope_atmodeller}a isolates the buffer contribution by shifting one module's converged offset by the analytic difference between the two iron-w\"ustite calibrations (the Hirschmann composite minus the $1\,\mathrm{bar}$ isoline of \citealt{Fischer2011EPSL304496F}), and this buffer-predicted curve nearly coincides with the \atmodeller\ curve in panel (a).
The residual gap in panel (b) therefore measures the difference between \CALLIOPE\ and \atmodeller\ as models, their solubility laws, equilibrium constants, and solver architectures, rather than their differing buffer calibrations, and is dominated by the sulfur solubility law.
The comparison spans $1800$ to $3000\,\mathrm{K}$, the surface-temperature range over which the crystallising magma oceans of this work spend nearly all of their evolution; at still higher temperatures silicate vapour species, inactive in both modules, contribute increasingly to the equilibrium \citep{Miguel2011ApJ742L19M,Zilinskas2022AA661A126Z,vanBuchem2023MPS581149V,vanBuchem2025AA695A154V,Seidler2026AA710A359S}, so a hotter comparison would probe that omitted chemistry rather than the module difference.
This agreement between two independently built chemical solvers underpins the oxidation-state results of Section~\ref{sec:results:redox_mode}.

\subsection{The coupled framework against the CHILI intercomparison}
\label{sec:val:main:chili}

\begin{figure*}
\centering
\includegraphics[width=\textwidth]{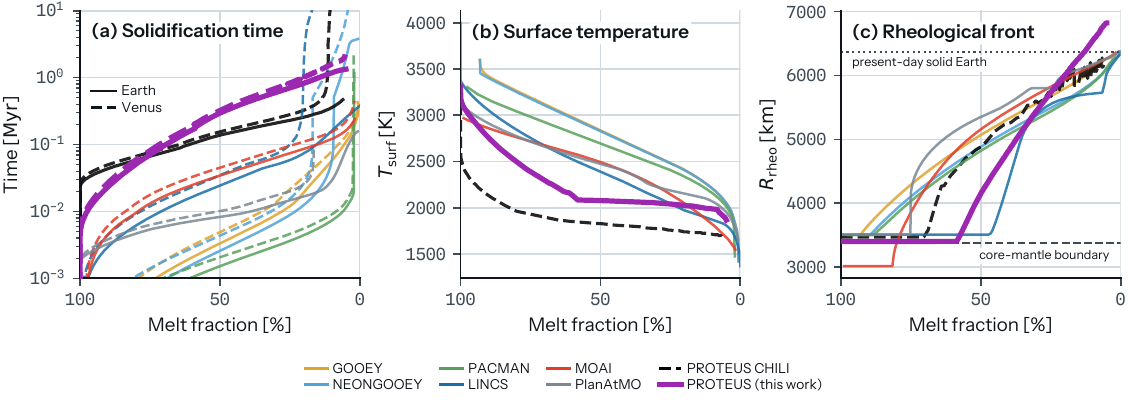}
\caption{
Validation of the coupled \PROTEUS\ framework against the community CHILI magma ocean intercomparison: the current \PROTEUS\ run (thick purple) against the six community models and the earlier submitted \PROTEUS-CHILI result (black dashed).
\textbf{(a)} Solidification time against melt fraction for the Nominal Earth (solid) and Nominal Venus (dashed) cases, so that all three panels share the melt-fraction axis; the current run reaches $5\,\%$ melt at $1.34\,\mathrm{Myr}$ (Earth) and about $5\,\%$ melt by $2.2\,\mathrm{Myr}$ (Venus), within the ensemble spread.
\textbf{(b)} Surface temperature and \textbf{(c)} rheological-front radius $R_\mathrm{rheo}$ against melt fraction for the Nominal Earth case, melt fraction decreasing left to right as the mantle solidifies; the present-day solid Earth radius and the \PROTEUS\ core-mantle boundary are marked in panel (c).
Curves show the global melt fraction; the community curves are volume-weighted (as was the \PROTEUS\ submission to the published intercomparison), while the current \PROTEUS\ curves are mass-weighted, and the two conventions agree to within a few per cent over this range.
}
\label{fig:val_chili}
\end{figure*}

The fully coupled framework reproduces the full community magma ocean intercomparison (Figure~\ref{fig:val_chili}).
Run in its full coupled configuration on the CHILI protocol \citep{Lichtenberg2026PSJ7108L}, whose primordial magma ocean intercomparison is published in \citet{Nicholls2026arXiv260624757N}, \PROTEUS\ advances the Nominal Earth and Nominal Venus magma oceans from their molten start through solidification, reaching a global melt fraction of $5\,\%$ at $1.34\,\mathrm{Myr}$ for Earth, within the $0.01$ to $2.0\,\mathrm{Myr}$ spread of the six independent community models (Figure~\ref{fig:val_chili}a).
Once the mantle is below $20\,\%$ melt by mass, the surface temperature lies within the envelope of surface temperatures defined by the community comparison, and the rheological-front radius tracks the ensemble to solidification (Figure~\ref{fig:val_chili}b,c).

The strongest departure of \PROTEUS\ from the community reference is at mid-crystallisation: once the mantle is $50\,\%$ solidified, the \PROTEUS\ surface temperature lies about $300\,\mathrm{K}$ below the coolest community model, so the atmosphere radiates less and the interior cools on the slow side of the ensemble, which the intercomparison attributes to its treatment of mantle dynamics and its melting curves \citep{Nicholls2026arXiv260624757N}.
The variance between models is driven chiefly by their treatments of volatile partitioning, notably into the solid mantle, and of atmospheric opacity \citep{Nicholls2026arXiv260624757N}, so the agreement is expected to loosen where those choices matter most.
The earlier \PROTEUS\ submission (black dashed line in Figure~\ref{fig:val_chili}), built on SPIDER with a static Adams-Williamson structure, shows an oscillatory rheological front; in the present framework the front migrates monotonically, a change we attribute to the $C^\infty$ regularisation of the phase-boundary and permeability factors (Section~\ref{sec:methods:aragog}), which lets the front move continuously rather than switching between discrete states.

The behaviour in Figure~\ref{fig:val_chili}c also foreshadows the central result of this paper.
The rheological front advances toward a radius larger than the present-day solid Earth because the interior structure in this reference configuration is computed once, at the hot molten start, and is not recontracted as the mantle cools, so the radius keeps the thermal expansion of the silicate melt.
Accounting for that melt-to-solid contraction, by recomputing the interior structure as the mantle crystallises rather than holding it fixed at the molten state, is exactly what the following section quantifies across the super-Earth mass range.

\section{Results}
\label{sec:results}

\subsection{Super-Earths shrink as they crystallise}
\label{sec:results:baseline}

\begin{figure}
\centering
\includegraphics[width=\columnwidth]{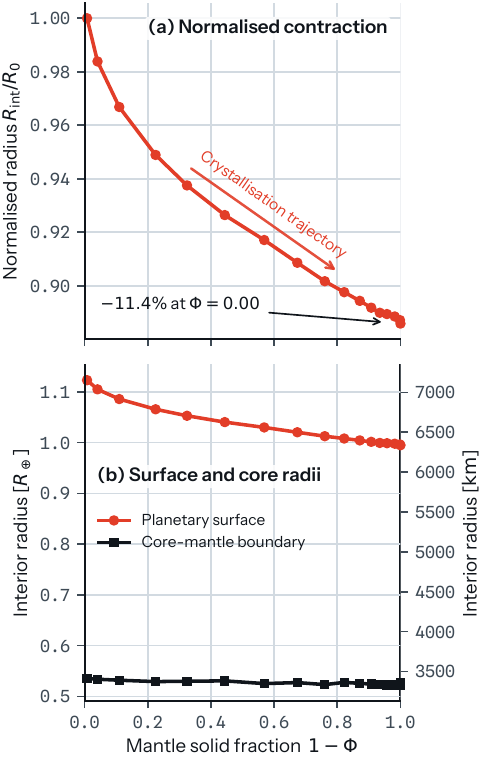}
\caption{Structural contraction of a crystallising super-Earth, from a standalone interior-structure sweep at $1\,\Mearth$ and an Earth-like core-mass fraction (PALEOS MgSiO$_3$ mantle on an adiabatic profile).
\textbf{(a)} Normalised interior radius $R_\mathrm{int}/R_0$ against mantle solid fraction $1-\Phi$; the radius contracts monotonically as the mantle crystallises, isolating the melt-to-solid density response from the cooling timeline and the atmosphere.
\textbf{(b)} Absolute interior radius $R_\mathrm{int}$ (planetary surface) and core radius $R_\mathrm{core}$ against mantle solid fraction, with the interior radius in km on the right axis; the contraction is produced by the thinning silicate mantle shell while the core radius is essentially fixed.
The melt fraction is mass-weighted over the mantle, and the reference radius $R_0$ of panel (a) is the interior radius of the molten starting state of the sweep.
\label{fig:shrinking}}
\end{figure}

\begin{figure*}
\centering
\includegraphics[width=\textwidth]{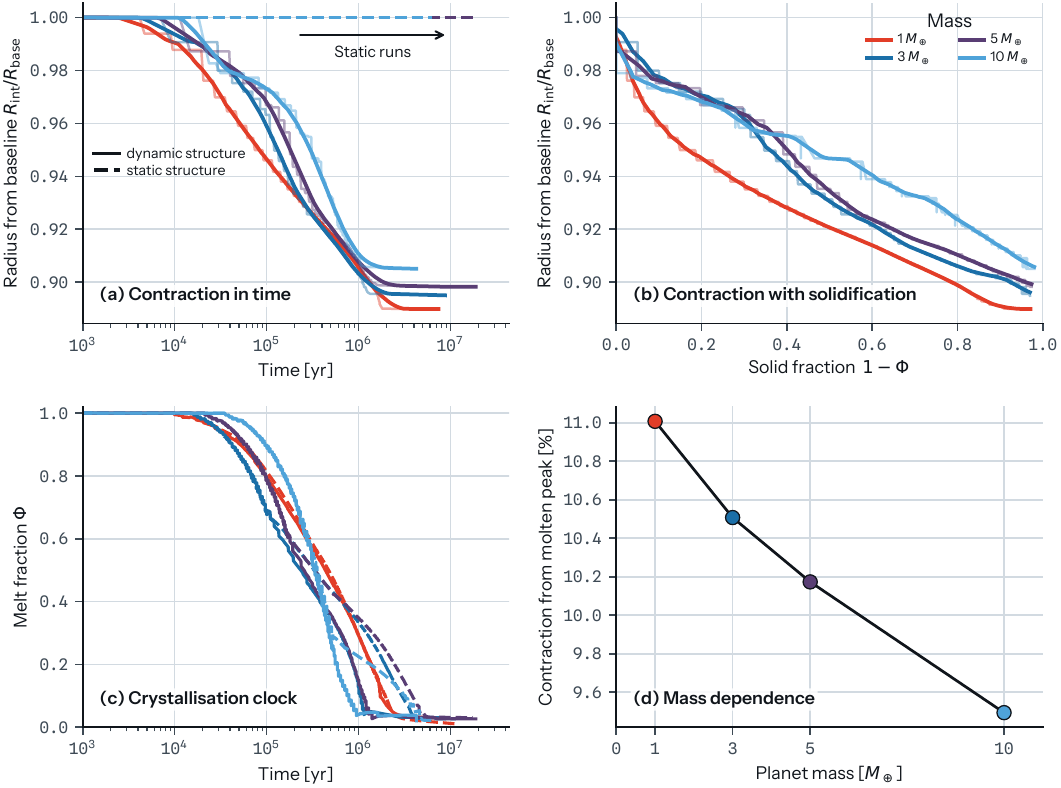}
\caption{Structural contraction across the simulation grid (planet mass $\{1, 3, 5, 10\}\,\Mearth$, dynamic versus static interior structure, IW$+4$, Sun at $1$\,AU), with the interior radius referenced to each mass's shared self-consistent baseline (the radius the dynamic and static runs share at the start, after which the structure of the static twin no longer evolves).
\textbf{(a)} Interior radius $R_\mathrm{int}/R_\mathrm{base}$ against time; the static twin (dashed) holds the baseline at unity while the dynamic run (solid) contracts below it as the mantle crystallises.
\textbf{(b)} Interior radius $R_\mathrm{int}/R_\mathrm{base}$ against solid fraction $1-\Phi$, dynamic runs only, collapsing the time axis; each track begins at the shared molten baseline, reached after the initial-condition transient.
In panels (a) and (b) the faint stepped curves are the unsmoothed dynamic-run output and the bold curves are moving averages of them: the interior structure is re-solved only when a temperature, melt-fraction, or composition threshold is crossed rather than at every timestep (Section~\ref{sec:methods:zalmoxis}), so the radius is held fixed between re-solves and the unsmoothed trace advances in visible steps.
\textbf{(c)} Global melt fraction $\Phi$ against time, dynamic (solid) versus static (dashed): the crystallisation history that panels (a) and (b) share.
\textbf{(d)} Total contraction from the molten start to the crystallisation endpoint against planet mass; the contraction decreases modestly with mass, from about $11\,\%$ at $1\,\Mearth$ to about $9.5\,\%$ at $10\,\Mearth$.
The magnitudes are measured at the crystallisation endpoints ($\Phi \approx 0.02$ to $0.03$) rather than at $\Phi = 0$, so they are close lower bounds on the fully solidified contraction; the decrease with mass is monotonic and robust.
The reference radius $R_\mathrm{base}$ is the self-consistent molten baseline that the dynamic and static runs of each mass share at the start.
\label{fig:contraction_grid}}
\end{figure*}

We find that the silicate mantle of a rocky planet contracts substantially as it cools and crystallises: thermal contraction of the superheated melt and the melt-to-solid density increase both reduce the interior radius at fixed mass and composition.
Both responses follow from the phase-aware PALEOS MgSiO$_3$ equation of state \citep{Attia2026subm}, in which the silicate liquid is more thermally expansive and more compressible than the solid, so the melt-to-solid density contrast that drives the contraction weakens toward the highest interior pressures; the full liquid and solid property tables are provided by \citet{Attia2026subm} and the accompanying data release rather than reproduced here.
Figure~\ref{fig:shrinking} isolates this structural contraction with a standalone interior-structure calculation at $1\,\Mearth$ and an Earth-like core-mass fraction, stepping the adiabat from a molten to a fully crystallised state and recording the interior radius against the mantle melt fraction $\Phi$.
The interior radius $R_\mathrm{int}$, normalised to its value $R_0$ in the molten starting state, falls monotonically as the mantle solid fraction $1-\Phi$ grows (Figure~\ref{fig:shrinking}a), contracting by about $11\,\%$ from the molten state ($\Phi = 0.99$) to a fully solid mantle ($\Phi = 0$).
In absolute terms the interior radius falls by about $810\,\mathrm{km}$, from about $7150$ to $6340\,\mathrm{km}$ (Figure~\ref{fig:shrinking}b, right axis).
The contraction is produced by the silicate mantle: its shell thins by roughly $20\,\%$ while the core radius decreases by only about $3\,\%$ (Figure~\ref{fig:shrinking}b).
The core and mantle exchange no mass in the model, so this small core contraction is the pressure response of the iron core to the thinning silicate shell above it rather than a transfer of material between them.
The fully crystallised interior recovers a radius of $0.99\,R_\oplus$, consistent with the cold rocky mass-radius relation at $1\,\Mearth$, so the contraction spans the difference between a hot, molten interior and its solidified end state.
Because this calculation prescribes the melt fraction and solves the structure for each value, rather than evolving the planet in time, it isolates the contraction from the cooling timeline and the atmosphere; the coupled dynamic-versus-static evolution runs (Figure~\ref{fig:contraction_grid}) confirm that the contraction of Figure~\ref{fig:shrinking} develops self-consistently and that holding the interior structure fixed removes it.

The grid of coupled, time-evolving models extends this contraction across the $1$ to $10\,\Mearth$ mass range and confirms that it develops self-consistently in time rather than from a prescribed melt fraction.
Throughout, contraction magnitudes are quoted at each run's crystallisation endpoint, the quasi-steady state in which the integration terminates (global melt fraction $\Phi \approx 0.02$ to $0.03$ for the runs of this subsection), and are therefore close to the fully solidified values: if the contraction stays proportional to the solid fraction over the last few per cent of melt, the quoted values fall short of the fully solidified ones by a few tenths of a percentage point.
The dynamic and static runs of each mass share a common self-consistent baseline radius at the start, against which the interior radius is referenced; the static twin holds that baseline while the dynamic run contracts below it as the mantle crystallises (Figure~\ref{fig:contraction_grid}a).
Expressed against solid fraction, the dynamic runs separate by mass into a family of contraction branches (Figure~\ref{fig:contraction_grid}b), the more massive interiors contracting less overall, while their static twins, with the interior structure held fixed at the shared baseline, keep the radius constant, so the difference between the two is the contraction.
The total contraction from the molten start to the crystallisation endpoint depends clearly on planet mass, decreasing from about $11\,\%$ at $1\,\Mearth$ to about $9.5\,\%$ at $10\,\Mearth$ (Figure~\ref{fig:contraction_grid}d), as a more strongly compressed massive interior returns a smaller fractional radius change for the same melt-to-solid density contrast.
At intermediate solid fractions the mass tracks are not strictly ordered (Figure~\ref{fig:contraction_grid}b), because each mass retains a different atmosphere that paces its cooling differently and the runs traverse the mush region on different timelines; the ordered mass dependence is a property of the crystallisation endpoints.
This variation in contraction of roughly $1.5$ percentage points is, however, small compared with the radius precision attainable from transit photometry, so the volatile-poor contraction that accompanies crystallisation can be summarised to first order as about $10\,\%$ across the $1$ to $10\,\Mearth$ range.

The degree of contraction described here is referenced to the molten start and therefore includes the thermal contraction of the superheated melt: in the coupled runs (fiducial superheat of $500\,\mathrm{K}$ above the liquidus) the interior radius falls by about $1$ to $2\,\%$ before crystallisation begins, and the remaining $8$ to $9$ percentage points accumulate as the mantle crystallises.
As Section~\ref{sec:results:stellar} shows, the precise values inherit the choice of initial superheat, whereas the solidified endpoint radius itself is invariant, set by mass and composition alone.
The contraction found here is roughly twice the $\sim5\,\%$ radius decrease that \citet{Bower2019AA631A103B} reported for a solidifying Earth-mass magma ocean.
The most likely origin is the silicate equation of state, since the contraction magnitude is set by the melt-to-solid density contrast, and the phase-aware PALEOS equation of state used here differs from that of \citet{Bower2019AA631A103B}.
The reference state may also contribute, since the value quoted here is measured from a superheated molten start and includes the pre-crystallisation thermal contraction noted above.
The contraction is a difference between two states of the same planet rather than a change that a survey would watch happen: the observable quantity is a radius excess at fixed mass relative to the cold mass-radius relation of the solidified end state.
Two degeneracies stand between that radius excess and the crystallisation state: the composition of the cold baseline, since the unknown iron fraction shifts the solidified radius by about $15\,\%$ across core-mass fractions of $0.2$ to $0.7$, more than the contraction signal, and also modulates the signal itself ($7.6$ to $10.7\,\%$ over the same range; Section~\ref{sec:results:stellar}), and the overlying atmosphere, which a transit radius does not separate from the interior.
Isolating the crystallisation signature therefore requires breaking both degeneracies jointly, and the most direct candidates are young planets caught before solidification, where the excess is largest; we develop this observational pathway in Sections~\ref{sec:disc:radii} and~\ref{sec:disc:obs}.

\subsection{Oxidation state and the free-oxygen budget}
\label{sec:results:redox_mode}

The oxidation state of the silicate melt sets the speciation of the outgassed atmosphere, and rocky planets are expected to span a wide range of mantle oxidation states, set by their accretion history, core formation, and interior chemistry \citep{Frost2008AREPS36389F,Sossi2020SciA61387S,Hirschmann2021GeCoA31374H}.
The coupled framework treats the oxidation state in two distinct ways (Section~\ref{sec:methods:atmodeller}).
In the fixed-fugacity treatment the surface oxygen fugacity is buffered at a prescribed offset from the iron-w\"ustite equilibrium throughout the evolution, so the oxidation state is imposed as a boundary condition.
In the oxygen-conserving treatment the free oxygen budget of the planet (Section~\ref{sec:methods:atmodeller}) is instead conserved as an elemental inventory alongside hydrogen, carbon, nitrogen, and sulfur, and the surface oxygen fugacity evolves freely as the magma ocean crystallises and the atmosphere grows.
Conserving the oxygen inventory in this way assumes that the free-oxygen abundance is otherwise invariant under the subsequent evolution, changing only through escape to space; this is a strong assumption, but so is the fixed oxygen fugacity that the buffered treatment imposes throughout.
We compare the two treatments on otherwise identical runs by taking the free oxygen budget calculated from each fixed-fugacity case and conserving this free-oxygen inventory in an oxygen-conserving twin during an evolutionary calculation.
We then report the trajectory of the derived fugacity offset and the free-oxygen budget, and test the equivalence of the two treatments at the initial outgassing equilibrium.

\begin{figure}
\centering
\includegraphics[width=\columnwidth]{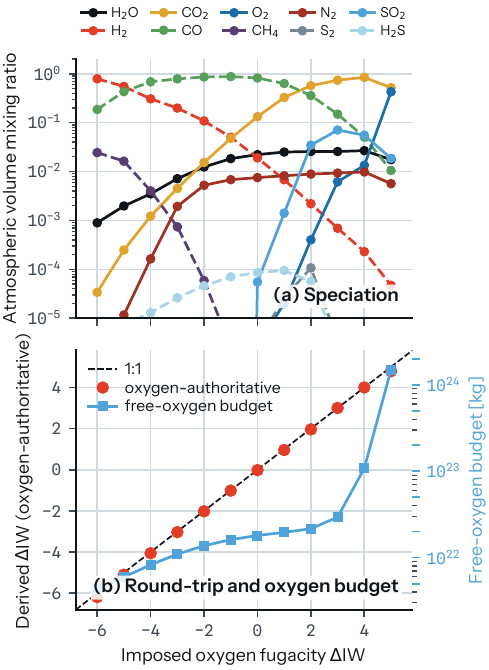}
\caption{Oxidation state and the free-oxygen budget of the fiducial $5\,\Mearth$ reference planet across the imposed oxygen-fugacity sweep (dynamic interior structure, Sun at $1$\,AU), comparing the fixed-fugacity and oxygen-conserving treatments.
\textbf{(a)} Atmospheric volume mixing ratios of the main C-H-O-N-S species in the fixed-fugacity treatment against the imposed iron-w\"ustite offset $\Delta\mathrm{IW}$, at a common crystallisation snapshot ($\Phi = 0.40$) so that the redox response is isolated from the differing crystallisation states of the runs.
\textbf{(b)} Equivalence of the two treatments and the free-oxygen budget against imposed $\Delta\mathrm{IW}$, sharing the horizontal axis.
Left axis: derived $\Delta\mathrm{IW}$ of the oxygen-conserving twins against the imposed $\Delta\mathrm{IW}$ of the fixed-fugacity twins, with the $1$:$1$ line; conserving the fixed-twin oxygen budget recovers the imposed oxygen fugacity to within a few hundredths of a log unit near iron-w\"ustite, loosening to about $0.2$\,dex at both extremes.
Right axis: the free (volatile-system) oxygen budget, comprising the atmospheric oxygen and the oxygen dissolved in the melt as volatile species, and excluding the mantle oxides, which increases monotonically and steepens toward the oxidising end, approaching the breakdown of the fixed-fugacity treatment at $\Delta\mathrm{IW}=+6$, where the equilibrium atmosphere would exceed the planet mass (excluded).
\label{fig:redox}}
\end{figure}

At a common crystallisation snapshot (Figure~\ref{fig:redox}a; $\Phi = 0.40$; surface temperatures of about $2450$ to $3000\,\mathrm{K}$ across the sweep) the outgassed atmosphere shifts systematically from reduced to oxidised across the fugacity range.
At the reducing end the atmosphere is dominated by molecular hydrogen, reaching a volume mixing ratio of about $0.79$ at $\Delta\mathrm{IW}=-6$, with carbon monoxide the second most abundant species; carbon monoxide overtakes hydrogen between $\Delta\mathrm{IW}=-5$ and $-4$ and then dominates the broad middle of the sweep, peaking near $0.88$ at $\Delta\mathrm{IW}=-1$ and remaining the most abundant species through iron-w\"ustite up to about $\Delta\mathrm{IW}=+1.5$.
Carbon dioxide rises steadily with $\fO$, overtakes carbon monoxide near $\Delta\mathrm{IW}=+1.5$, and dominates the oxidising branch, reaching a volume mixing ratio of about $0.84$ at $\Delta\mathrm{IW}=+4$.
In the scenario buffered at fixed $\Delta\mathrm{IW}$, molecular oxygen stays subdominant across the sweep, rising steeply only at the most oxidising end to reach a volume mixing ratio of about $0.43$, comparable to carbon dioxide, at $\Delta\mathrm{IW}=+5$.
Sulfur tracks the same reduced-to-oxidised transition, shifting from H$_2$S at the reducing end to SO$_2$ at the oxidising end, where SO$_2$ peaks near $7\,\%$ by volume at $\Delta\mathrm{IW}=+3$.
Water vapour remains a minor constituent throughout, so the transition is driven by the carbon, sulfur, and free-oxygen system rather than by water.
This reduced-to-oxidised speciation sequence follows the same oxygen-fugacity ordering found for Hadean-Earth outgassing models \citep{Sossi2020SciA61387S,Gaillard2022EPSL57717255G}.
Under the same buffer the surface pressure rises gently from about $260\,\mathrm{bar}$ at the reducing end to about $500\,\mathrm{bar}$ near iron-w\"ustite, where the redox state mainly redistributes a fixed volatile mass among species of comparable mean molecular weight, and increases modestly toward the oxidising end, from about $500\,\mathrm{bar}$ at $\Delta\mathrm{IW}=0$ to about $860\,\mathrm{bar}$ at $\Delta\mathrm{IW}=+5$, so the surface pressure of the buffered atmosphere varies by only a factor of about three across the sweep.
The compositional shift links the oxidation state to the planetary radius: across the sweep of Figure~\ref{fig:redox}a the mean molecular weight of the outgassed atmosphere rises monotonically from about $7\,\mathrm{g\,mol^{-1}}$ at the hydrogen-dominated reducing end to about $42\,\mathrm{g\,mol^{-1}}$ at the carbon dioxide-dominated oxidising end, compressing the atmospheric scale height by a factor of about six at fixed temperature and gravity.
The oxidation state thus shapes both the atmosphere's contribution to the transit radius and its spectroscopic accessibility.

The oxygen-conserving twins recover the imposed oxygen fugacity at equal oxygen (Figure~\ref{fig:redox}b).
Conserving the free oxygen budget calculated from each fixed-fugacity run and deriving $\fO$ as a property of the system, we find that the derived $\Delta\mathrm{IW}$ tracks the imposed value to within a few hundredths of a log unit across most of the range, confirming the equivalence of the two treatments (Section~\ref{sec:methods:atmodeller}) on the full $\Delta\mathrm{IW}$ sweep rather than at a single point.
The agreement is tightest near iron-w\"ustite, where $\Delta\mathrm{IW}=0$ is recovered as $-0.02$, and loosens at the extremes, where the oxygen closure of the initial equilibrium degrades to about $5$ to $6\,\%$ of the budget and the derived offset reaches $-6.2$ at $\Delta\mathrm{IW}=-6$ and $+4.8$ at $\Delta\mathrm{IW}=+5$.
The conserved oxygen inventory increases monotonically with the imposed fugacity (Figure~\ref{fig:redox}b, right axis), from about $4\times10^{21}\,\mathrm{kg}$ at $\Delta\mathrm{IW}=-6$ to $1.5\times10^{24}\,\mathrm{kg}$ at $\Delta\mathrm{IW}=+5$, and steepens sharply at the oxidising end, rising by more than an order of magnitude over the final two fugacity steps.
This monotonicity is what makes the oxygen-conserving closure well posed (Section~\ref{sec:methods:atmodeller}): each feasible oxygen target corresponds to exactly one fugacity offset.
Table~\ref{tab:oxygen_budget} lists this conserved oxygen budget and its ratio to the fixed hydrogen, carbon, nitrogen, and sulfur inventory at each fugacity level.

The two treatments encode the same oxidation state at the initial outgassing equilibrium (Figure~\ref{fig:redox}b), the equivalence tested here.
As the mantle crystallises the conserved-oxygen treatment can build a more oxygen-rich atmosphere than the buffered treatment at the most oxidising end; we do not analyse that regime in detail, since the outgassed radiative transfer does not yet include molecular-oxygen opacity.
At $\Delta\mathrm{IW}=+6$ the initial outgassing equilibrium demands an oxygen-bearing atmosphere whose mass exceeds that of the planet, so this degree of oxidation is excluded and the physical sweep spans $\Delta\mathrm{IW}=-6$ to $+5$.
The redox and speciation results reported above follow from the outgassing equilibrium and the conserved oxygen budget and are independent of the cooling rate, which the atmosphere sets through the surface interior flux $F_\mathrm{int}$.
At the initial outgassing equilibrium of the fiducial $\Delta\mathrm{IW}=+4$ reference, the outgassed atmosphere is only weakly sensitive to the adopted sulfur budget, which is set well below the bulk-silicate-Earth value (Section~\ref{sec:methods:anchors}).
Raising the sulfur budget twentyfold, to the bulk-silicate-Earth value, increases the SO$_2$ abundance roughly in proportion but changes the surface pressure by only about $3\,\%$, leaving the oxygen-dominated atmosphere otherwise unchanged.

\subsection{Volatile inventory and the high-pressure regime}
\label{sec:results:volatiles}

\begin{figure}
\centering
\includegraphics[width=\columnwidth]{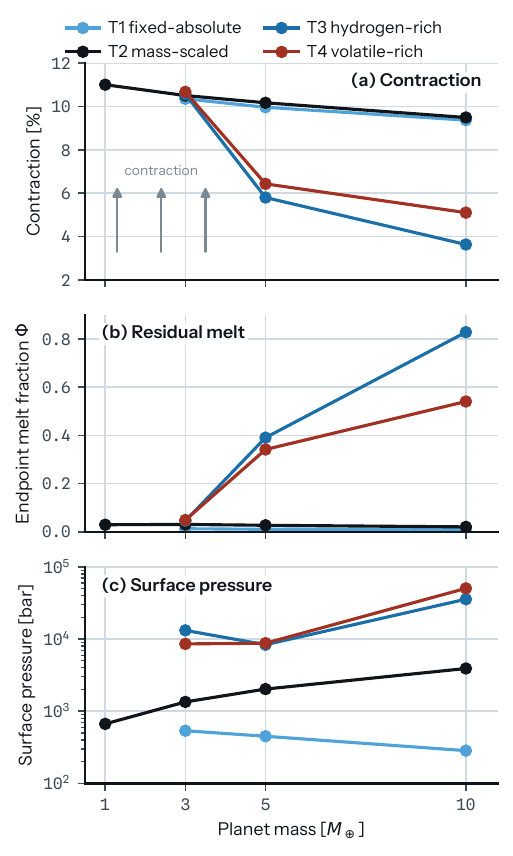}
\caption{Volatile inventory and the high-pressure regime across planet mass, for the four volatile-inventory treatments of Table~\ref{tab:volatile_treatments} at fixed oxidation state ($\Delta\mathrm{IW}=+4$) and dynamic interior structure.
\textbf{(a)} Total crystallisation contraction, measured from the molten peak, against planet mass; the volatile-poor (fixed-absolute) and mass-scaled fiducial interiors follow the contraction of Section~\ref{sec:results:baseline}, while the hydrogen-enriched and volatile-rich interiors contract far less at high mass.
The upward arrows mark the direction of increasing crystallisation contraction: a model climbs this axis as it contracts, the opposite sense to the interior-radius panels of Figures~\ref{fig:shrinking} and~\ref{fig:contraction_grid}, where a contracting model descends.
\textbf{(b)} Endpoint melt fraction against mass; the volatile-poor and mass-scaled cases solidify to $\Phi \approx 0.01$ to $0.03$ at all masses, whereas the hydrogen-enriched and volatile-rich cases retain $\Phi$ up to $0.83$ at $10\,\Mearth$, a deep magma ocean sustained at radiative equilibrium.
\textbf{(c)} Surface pressure at the crystallisation endpoint against mass (logarithmic); the volatile ladder spans a few hundred bar to about $5\times10^{4}\,\mathrm{bar}$, with the thickest atmospheres on the most volatile-rich, most massive cases.
The fixed-absolute inventory (a mass-independent volatile mass) thins with mass while the mass-scaled inventory grows with it.
\label{fig:volatiles}}
\end{figure}

The volatile inventory sets the mass and composition of the outgassed atmosphere, and through it the rate at which the mantle can radiate its heat and crystallise.
We investigate these volatile regimes in the simulations with the four treatments of Table~\ref{tab:volatile_treatments}, the volatile-poor fixed-absolute case, the mass-scaled fiducial inventory, the hydrogen-enriched case, and the volatile-rich case, at fixed oxidation state ($\Delta\mathrm{IW}=+4$) and dynamic interior structure, across the $1$ to $10\,\Mearth$ mass range (Figure~\ref{fig:volatiles}); the mass-scaled fiducial is the case we explored in Section~\ref{sec:results:baseline}.
The treatments produce surface pressures at the crystallisation endpoint that range from a few hundred bar for the volatile-poor planets up to about $5\times10^{4}\,\mathrm{bar}$ for the volatile-rich and hydrogen-enriched planets (Figure~\ref{fig:volatiles}c).
The mass-scaled volatile inventory grows the atmosphere with the planet, from about $670\,\mathrm{bar}$ at $1\,\Mearth$ to about $3900\,\mathrm{bar}$ at $10\,\Mearth$, whereas the fixed-absolute volatile inventory, held at a mass-independent volatile mass, thins with increasing mass as the same volatile mass is spread over a larger surface area.

The volatile inventory sets a threshold above which the interior no longer solidifies and the structural contraction stalls.
At low mass every treatment crystallises to its endpoint (global melt fraction $\Phi \approx 0.01$ to $0.05$) and its interior contracts by about $10$ to $11\,\%$, indistinguishable from the mass-scaled baseline (Figure~\ref{fig:volatiles}a,b).
At higher mass, however, the hydrogen-enriched and volatile-rich planets fail to solidify: their thick atmospheres throttle the surface heat loss so strongly that the interior settles into a deep magma ocean at radiative equilibrium \citep{Hamano2013Natur497607H}, retaining a global melt fraction between $0.34$ and $0.83$ across the $5$ and $10\,\Mearth$ cases.
These interiors therefore reach a radiative quasi-steady state rather than solidifying, and this is the criterion on which the integration is stopped: the run terminates once the net planetary energy imbalance falls below an absolute tolerance of $1\,\mathrm{W\,m^{-2}}$ (or a relative tolerance of $10^{-3}$), at which point the interior heat loss has effectively stalled and the melt fraction plateaus rather than crystallising further within the $6\,\mathrm{Gyr}$ simulation runtime.
Because the contraction is realised only as the mantle crystallises (Section~\ref{sec:results:baseline}), these incompletely solidified interiors contract by only about $3.5$ to $6.5\,\%$, well short of the about $10\,\%$ reached by their volatile-poor counterparts of the same mass.
The volatile inventory therefore sets not only the atmospheric mass and composition, but whether the crystallisation-driven contraction of the interior is realised at all: a volatile-rich super-Earth can remain inflated by a persistent magma ocean that its own atmosphere sustains, a distinct route to a radius excess from the mid-crystallisation state of Section~\ref{sec:results:baseline}.
The mass at which this transition appears follows from the mass-volatile scaling adopted here (Table~\ref{tab:volatile_treatments}) rather than marking a fundamental threshold: the volatile inventory is the controlling quantity, so a more volatile-rich planet reaches the non-solidifying regime at a lower mass, whereas a planet that stays volatile-poor can crystallise fully at a higher mass.
Such a body, a $5$ to $10\,\Mearth$ interior bearing a thick hydrogen-rich atmosphere over a persistent magma ocean (model photospheric radii of about $1.6$ to $2.0\,\Rearth$, subject to the dry-interior simplification of Section~\ref{sec:methods:anchors}), would conventionally be classed as a sub-Neptune rather than a super-Earth; where the physically or chemically defining boundary between the two lies in this mass regime is, however, neither sharp nor settled and remains debated \citep{Lichtenberg2025Sci390S3660L,Madhusudhan2025PNAS12216194M}.

The most volatile-rich, most massive interiors have the thickest atmospheres in the grid, with surface pressures reaching about $5\times10^{4}\,\mathrm{bar}$ (Figure~\ref{fig:volatiles}c).
These pressures remain below the $10^{5}\,\mathrm{bar}$ validity limit of the ideal-gas outgassing thermodynamics (Section~\ref{sec:methods:anchors}).
It is this thick atmosphere that suppresses the surface heat loss and sustains the deep magma ocean, so the same volatile enrichment that thickens the atmosphere is what keeps the interior molten and the planet inflated; the same thick atmosphere also dominates what a transit measures, an overlap we investigate in Section~\ref{sec:disc:radii}.
Whether, and on what timescale, such an interior eventually solidifies depends sensitively on the initial volatile budget and on the long-term atmosphere-loss history; deriving these timescales is not a goal of this work, and we refer to coupled evolution studies that quantify them \citep{Lichtenberg2021JGRE12606711L,Nicholls2024JGRE12908576N,Nicholls2026NatAs10809N,Sastre2026arXiv260620249S,Postolec2026arXiv260715011P,Calder2026MNRAS549g1007C}.

\subsection{Stellar environment and structural robustness}
\label{sec:results:stellar}

\begin{figure}
\centering
\includegraphics[width=\columnwidth]{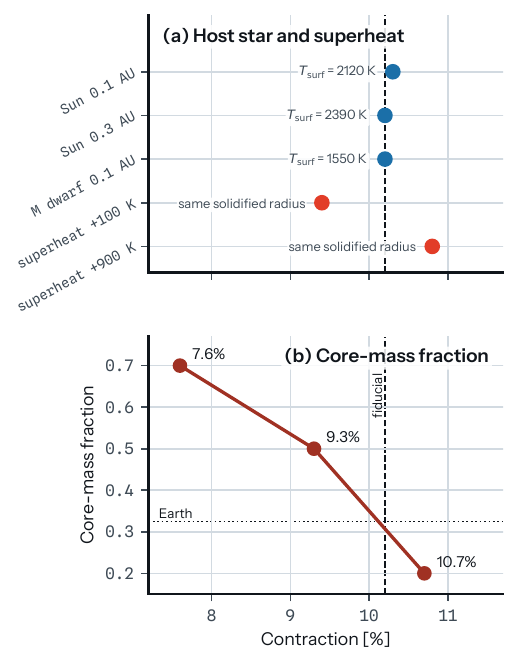}
\caption{Structural robustness of the crystallisation contraction across single-parameter variations around the fiducial $5\,\Mearth$ reference.
\textbf{(a)} Total contraction for the host-star and irradiation variations (blue) and the initial-superheat variations (amber), against the fiducial value (dashed); the host star and the instellation leave the contraction unchanged, setting the endpoint surface temperature (annotated) rather than the contraction, and the two superheat starts reach the same solidified radius, their apparent contraction differing only because it is referenced to the more expanded hot-start peak.
\textbf{(b)} Total crystallisation contraction against core-mass fraction; the contraction is produced by the silicate shell, so a more iron-rich planet, with less mantle to crystallise, contracts less.
The magnitudes are measured at the crystallisation endpoints (Section~\ref{sec:results:baseline}).
\label{fig:robustness}}
\end{figure}

The contraction of Section~\ref{sec:results:baseline} is a property of the cooling and crystallising silicate mantle, and a set of single-parameter variations around the fiducial $5\,\Mearth$ reference (Figure~\ref{fig:robustness}) establishes what the planet's stellar environment, structure, and initial thermal state each control, and what leaves the contraction unchanged.
Each variation holds the reference setup fixed and steps one parameter, so the contraction, measured from the molten peak as in Section~\ref{sec:results:baseline}, is directly comparable across the set.

Varying the host star and the flux received by the planet leaves the contraction essentially unchanged.
At $5\,\Mearth$ the total crystallisation contraction is $10.3\,\%$ for a planet at $0.1\,\mathrm{AU}$ around a Sun-like star, $10.2\,\%$ at $0.3\,\mathrm{AU}$ around the same star, and $10.2\,\%$ at $0.1\,\mathrm{AU}$ around a $0.2\,M_\odot$ M dwarf (Figure~\ref{fig:robustness}a), a spread of about a tenth of a percentage point across a change of spectral type and more than an order of magnitude in bolometric instellation.
The structural contraction is set by the interior rather than by the radiation field, for planets that do solidify, as expected when the atmosphere only paces the crystallisation.
What the planet's environment sets instead is the cooling timeline and the residual surface state: the equilibrium surface temperature at the crystallisation endpoint is set by the atmosphere the planet retains rather than by the incident flux directly.
At the same $0.1\,\mathrm{AU}$ the endpoint surface temperature settles near $1550\,\mathrm{K}$ for the M dwarf against $2120\,\mathrm{K}$ for the more luminous Sun-like star (bare equilibrium temperatures of about $280$ and $910\,\mathrm{K}$), a far smaller contrast than the order-of-magnitude instellation difference would imply, and placing the planet at $0.3\,\mathrm{AU}$ around the same Sun-like star raises the endpoint temperature to about $2390\,\mathrm{K}$ rather than lowering it.
This inversion follows from atmospheric escape \citep{Postolec2026arXiv260715011P}: the more strongly irradiated close-in planet loses roughly an order of magnitude more of its atmosphere, so it reaches the crystallisation endpoint with a thinner atmosphere and hence a cooler surface, the temperatures annotated in Figure~\ref{fig:robustness}a.
The time to reach the endpoint likewise differs between the three runs, set by the net flux that the retained atmosphere and the instellation together allow the interior to lose.
The contraction itself, however, is insensitive to these differences, so our conclusion that super-Earth interiors contract by about $10\,\%$ is unaffected by these atmospheric processes and holds across the close-in, strongly irradiated regime that dominates the detectable super-Earth population.

The core-mass fraction sets the magnitude of the contraction through the size of the silicate shell that crystallises.
Increasing the core mass fraction from $0.2$ to $0.5$ to $0.7$ at fixed mass reduces the contraction from $10.7$ to $9.3$ to $7.6\,\%$ (Figure~\ref{fig:robustness}b), while the core radius grows from about $0.42$ to $0.73$ of the interior radius.
This is the same partition seen in the reference planet (Section~\ref{sec:results:baseline}), where the contraction is produced by the thinning mantle and the core radius is nearly fixed: a more iron-rich planet has less silicate to crystallise, so the same melt-to-solid density contrast produces a smaller fractional change in the interior radius.
The contraction is therefore a mantle property, scaling with the silicate fraction.

In these models the solidified radius endpoint loses its memory of the initial thermal state.
Starting the reference planet from a strongly superheated melt ($900\,\mathrm{K}$ above the liquidus, against $500\,\mathrm{K}$ in the fiducial run) rather than a mildly superheated one ($100\,\mathrm{K}$) changes the molten starting radius but not the crystallised one: the two runs reach nearly the same final interior radius ($9853$ against $9846\,\mathrm{km}$) and surface temperature ($1790$ against $1750\,\mathrm{K}$), and their apparent contraction differs ($10.8$ against $9.4\,\%$) only because the hot start first removes about $2\,\%$ of its radius by thermal contraction of the fully molten mantle before crystallisation begins.
The crystallised structure is thus set by the mass and composition, independent of how much heat the planet began with, which is the basis for comparing runs that start from a common super-liquidus condition.

This set of variations, taken together, shows the contraction to be a stable signature of the cooling and crystallising silicate interior: fixed by the mantle and its iron fraction and indifferent to the host star, while the initial superheat shifts its magnitude by less than 1 percentage point and leaves the solidified radius unchanged.
As in Section~\ref{sec:results:baseline}, these magnitudes are measured at the crystallisation endpoints rather than at full solidification, so they are close lower bounds; the trends reported here are robust to the last few per cent of melt.

\section{Discussion}
\label{sec:discussion}

\subsection{Interpreting the radii of super-Earths}
\label{sec:disc:radii}

The crystallisation contraction established in Section~\ref{sec:results:baseline} is a stable, mantle-set signature of the cooling interior rather than an artefact of the cooling timeline or the atmosphere, and it changes how a measured radius is interpreted.
A super-Earth that retains a hot, partially molten interior is larger than its fully solidified counterpart of the same mass and composition, so a radius offset above the cold mass-radius relation encodes the thermal and crystallisation state of the mantle rather than only its bulk composition.
Secular contraction of a cooling rocky body is observed directly in the Solar System: Mercury's lobate scarps and wrinkle ridges record a global radial contraction of up to about $7\,\mathrm{km}$ of its solid mantle and core \citep{Byrne2014NatGe7301B}.
That contraction is far smaller than the crystallisation contraction found here, but it is the same kind of cooling-driven volume change.
Because the contraction scales with the silicate fraction and is nearly independent of the radiation environment and only weakly dependent on the initial superheat (Section~\ref{sec:results:stellar}), a measured radius excess at fixed mass constrains the crystallisation state of the rocky mantle.
This is a handle on interior state complementary to the compositional degeneracy that dominates static mass-radius interpretation \citep{Dorn2015AA577A83D,Unterborn2018NatAs2297U,Zeng2019PNAS1169723Z}, and it aligns with recent efforts to infer interior properties from coupled interior-atmosphere observables \citep{Lichtenberg2025Sci390S3660L}.
A radius alone, however, does not separate the interior from the atmosphere: a transit radius conflates the solid or molten mantle with any overlying volatile envelope \citep{Nicholls2026arXiv260415891N}, and disentangling the two demands spectral characterisation of the atmosphere rather than the bulk density alone.
This conflation is sharpest exactly where the interior radius excess is largest: the planets whose interiors stay inflated by a persistent magma ocean (Section~\ref{sec:results:volatiles}) are those bearing the thickest atmospheres in the grid.
The atmospheric and interior contributions to their transit radius therefore overprint each other, and separating the inflated-interior population requires atmospheric characterisation alongside the radius.
The contraction quantified here further assumes a volatile-free silicate interior (Section~\ref{sec:disc:limitations}), a simplification that matters because the dissolved-volatile contribution to the radius is itself likely important to the radius separation seen in the low-mass exoplanet census, the radius valley \citep{Fulton2017AJ154109F,VanEylen2018MNRAS4794786V,Owen2017ApJ84729O,Burn2024NatAs8463B,Heng2025ApJ99428H}.
Dissolved volatiles inflate the silicate interior and, at a fixed total inventory, shrink the total radius by drawing mass from the atmosphere into the melt (Attia et al., in preparation), an exchange the interior structure calculation here does not include.

\subsection{Molten interiors across the low-mass census}
\label{sec:disc:molten}

The interior phase state is a critical quantity to constrain, since a mantle kept molten by processes other than its own residual heat shows a radius excess, as does a young, still-cooling one; tidal heating in particular can sustain a magma ocean over gigayear timescales in close-in and eccentric systems \citep{Farhat2025ApJ979133F,Nicholls2025MNRAS5412566N,Seligman2024ApJ96122S,Peterson2023Natur617701P}, decoupling the radius offset from the planet's cooling age and driving dynamical variability in the melt itself \citep{Farhat2026ApJ1003208F}.
More broadly, the crystallisation state of the mantle is a largely unexplored axis in the interpretation of the low-mass census.
The rocky and water-rich interpretations of the small-planet population depend on bulk density at fixed mass \citep{Luque2022Sci3771211L,Venturini2024AA686L9V,Burn2024NatAs8463B}, a division that remains contested \citep{Rogers2023ApJ947L19R,Parc2024AA688A59P}; part of the density spread now attributed to composition may instead record the thermal and phase state of the interior.
This matters all the more because a growing body of work argues that many low-mass planets retain magma oceans rather than fully solidifying, as molten rocky sub-Neptunes \citep{Calder2026MNRAS549g1007C,Bean2021JGRE12606639B,Kite2019ApJ887L33K}, an interpretation now extended to individual planets whose JWST spectra are explained by a magma-ocean interior or its cloudy atmosphere-mantle interface \citep{Shorttle2024ApJ962L8S,Nixon2025ApJ99595N,Mukherjee2026ApJ1005L63M}.
This body of work adds to the classic result that sufficiently irradiated or volatile-rich worlds resist solidification altogether \citep{Hamano2013Natur497607H} and to our own volatile-rich cases that hold a deep magma ocean at radiative equilibrium (Section~\ref{sec:results:volatiles}).
Taken together, these results suggest that the cold, solidified end state assumed by the static structure models in common use, and by the interior retrievals built on them, may be the exception rather than the rule across the low-mass planet population.
Whether a departure of that kind leaves a statistically detectable imprint on the observed population is the sort of question survey-level demographic tests are built to answer, as they are for the runaway-greenhouse transition expected at the inner edge of the habitable zone \citep{Schlecker2024PSJ53S}.

\subsection{Observational prospects}
\label{sec:disc:obs}

These distinctions in radius and atmosphere are becoming observationally accessible in the close-in and ultra-short-period rocky planets now targeted by JWST, where the interior phase state and the presence of an atmosphere are jointly in question.
Ultra-short-period planets are the most amenable to this characterisation through their strong dayside thermal emission, and their outgassed atmospheres reflect the oxidation state and volatile budget of the interior below.
Secondary-eclipse and phase-curve measurements are being used to decide whether such planets are bare rocks or retain atmospheres \citep{Kreidberg2019Natur57387K,Greene2023Natur61839G,Zieba2023Natur620746Z,Gillon2026NatAs10674G,Crossfield2022ApJ937L17C,MeierValdes2025AA698A68M}, with a secondary atmosphere detected on the lava world 55~Cancri~e \citep{Hu2024Natur630609H,Patel2024AA690A159P,Snellen2025ARAA6383S}, a thick volatile envelope inferred for the ultra-hot super-Earth TOI-561~b \citep{Teske2025ApJ995L39T}, and rock-vapour mineral atmospheres expected at the hottest surfaces \citep{Miguel2011ApJ742L19M,Zilinskas2022AA661A126Z,Zilinskas2023AA671A138Z,Seidler2026AA710A359S}.
Because a transit radius alone cannot separate a molten interior from an overlying atmosphere, this discrimination requires exactly the thermal and spectral characterisation these programmes are built to provide \citep{Allen2026AJ171105A,Xue2025ApJ995L52X,TRAPPIST1JWSTCommunityInitiative2024NatAs8810T}, and interpreting a radius excess as a signature of incomplete crystallisation will demand the same joint interior-atmosphere inference extended across the population \citep{Lichtenberg2025TrGeo751L,Coy2025ApJ98722C,Apai2025PSJ6165A}.
A particularly direct route to a molten interior is offered by the youngest transiting planets, now being discovered and characterised in nearby young clusters and moving groups \citep{Shivkumar2026arXiv260701559S}, where the interior is most likely to be caught in or near its magma-ocean stage \citep{Lupu2014ApJ78427L,Bonati2019AA621A125B,Cesario2024AA692A172C,Cesario2026arXiv260713793C,Kimura2026ApJ1000220K,Panagiotou2026arXiv260715204P}.
JWST is already resolving the atmospheres of such young sub-Neptunes and their probable rocky progenitors \citep{Barat2024NatAs8899B,Barat2024AA692A198B,Barat2025AJ170165B,Barat2026ApJ1002L32B,Murphy2026AJ17266M}, several inferred to retain hot interiors.
The fraction of the observed population in which such a radius excess is detectable is not quantified here and is left to a dedicated demographic analysis.

\subsection{Oxidation state and atmospheric composition}
\label{sec:disc:redox}

The oxidation-state results (Section~\ref{sec:results:redox_mode}) show that whether $\fO$ is prescribed or derived, not only its value, shapes the outgassed atmosphere.
The speciation shifts systematically from a hydrogen-and-carbon-monoxide atmosphere at the reducing end through carbon dioxide at the oxidising end, with sulfur shifting from hydrogen sulfide to sulfur dioxide, and the conserved free-oxygen budget increases monotonically with the imposed fugacity \citep{Sossi2020SciA61387S,Bower2022PSJ393B,Suer2023FrEaS1159412S,Nicholls2024JGRE12908576N}.
The fixed-fugacity prescription standard in earlier work breaks down at the oxidising extreme, where it demands an atmosphere more massive than the planet, whereas the oxygen-conserving treatment introduced here stays well posed because the oxygen inventory is finite and conserved.
The two formulations encode the same oxidation state at the initial equilibrium, but for the most oxidised planets they diverge as crystallisation proceeds, because only the conserved-oxygen treatment can develop an abiotic molecular-oxygen atmosphere.
The oxygen-conserving mode is therefore the appropriate tool for the oxidised corner most relevant to the interpretation of abiotic molecular oxygen on rocky exoplanets \citep{Schaefer2016ApJ82963S,Wordsworth2014ApJ785L20W,Luger2015AsBio15119L,KrissansenTotton2021AGUA200294K}.

\subsection{Limitations}
\label{sec:disc:limitations}

A few limitations bound our results.
The crystallisation contractions are measured at the crystallisation endpoints (Section~\ref{sec:results:baseline}), where most interiors reach global melt fractions of a few per cent, rather than at complete solidification, so they are close lower bounds on the fully solid contraction; the most volatile-rich and most oxidised cases instead hold a deep magma ocean at radiative equilibrium and are discussed as such.
The cooling timescales are compared in relative terms across the grid rather than as absolute clocks, since the absolute timescale depends on the interior energy balance.
For the volatile-rich planets that do not solidify, the epoch at which the contraction would eventually be realised is set by how long the magma ocean persists, which follows in turn from the volatile inventory the planet begins with and from how much of that inventory is lost to escape; putting a number on that timeline lies outside the scope of this work and is the subject of dedicated coupled evolution models \citep{Lichtenberg2021JGRE12606711L,Nicholls2024JGRE12908576N,Nicholls2026NatAs10809N,Sastre2026arXiv260620249S,Postolec2026arXiv260715011P,Calder2026MNRAS549g1007C,Cesario2026arXiv260713793C}.
The contraction also feeds back on the convective dynamics of the magma ocean itself: as the silicate shell thins, its aspect ratio and the mixing length that sets the convective diffusivity (Section~\ref{sec:methods:aragog:eddy}) shrink with it, a feedback the dynamic structure updates capture to first order, although for strongly iron-rich planets with thin shells the convective regime itself may shift in ways that mixing-length theory does not resolve.
The outgassing thermodynamics, including the cross-module comparison of Section~\ref{sec:framework_validation}, use the ideal-gas closure throughout.
Real-gas fugacity corrections at surface pressures of $10^4$ to $10^5$~bar change fugacities, speciation, and element partitioning, and can even reverse the sign of the radius difference between volatile-rich and volatile-poor planets (Attia et al., in preparation).
The deep-magma-ocean endpoint radius of the most volatile-rich cases therefore depends on the atmospheric thickness and is correspondingly uncertain, and incorporating these corrections is future work.
At the oxidising corner the radiative transfer omits molecular-oxygen opacity, so the surface fluxes there are upper bounds, and the oxygen-conserving surface pressures above an iron-w\"ustite offset of $+4$ exceed the validity limit of the outgassing thermodynamics and are extrapolations.
The evolution is one-dimensional.
The crystallisation contraction is a global volume change set by the planet mass, composition, and phase state, which a radial average captures; a one-dimensional model omits lateral structure, such as convective cells, a latitudinal cooling contrast, or a hemispheric magma ocean on a tidally locked planet \citep{Meier2023AA678A29M,Boukare2025NatAs91511Ba,Meier2026MNRAS547ag390M}, which redistributes where heat is lost.
The interior uses a single-component MgSiO$_3$ equation of state for both liquid and solid: the mantle contains no iron or other cations apart from Mg and Si, so there is no chemical fractionation between solid and melt and no iron enrichment of the residual melt during progressive solidification.
An iron-enriched melt would be denser and could reverse the liquid-solid density contrast at depth, weakening the one-way draining assumed here, and the residual melt would then remain distributed by the mush criterion rather than forming a separate basal reservoir.
The interior structure calculation (i.e., the equation of state) includes no dissolved volatiles: dissolving them into the melt would raise the interior radius while, at fixed total inventory, contracting the total radius through the exchange of mass from atmosphere to interior \citep{Dorn2021ApJ922L4D}, and neither effect enters the structure solve here.
Atmospheric escape is treated with a single parameterisation whose long-term mass loss is uncertain, and we do not survey the range of escape histories that would reshape the volatile inventory and the surface state \citep{Owen2019AREPS4767O,Wordsworth2022ARAA60159W,Lammer2008SSRv139399L}.
The equation of state treats silicate and volatiles as immiscible, whereas hydrogen and silicate can mix at the pressures and temperatures of massive molten interiors, blurring the interior-atmosphere boundary and altering the radius in a way that is distinct from, though observationally degenerate with, the crystallisation contraction quantified here \citep{Young2024PSJ5268Y,Young2025PSJ6251Y,Rogers2025MNRAS5443496R,Rogers2026arXiv260630644R,Steinmeyer2026ApJ100136S}.
The volatile inventory is prescribed rather than grown, so the accretion and formation pathways that set it, and the diversity they produce, are not modelled \citep{Kimura2022NatAs61296K,Kimura2026ApJ1000220K,Venturini2020AA643L1V,Miozzi2025Natur648551M}.
Finally, we do not confront the crystallisation radius signature with a specific observed system here; connecting it to individual measured planets is left to future work.

\section{Conclusions}
\label{sec:conclusions}

This work makes two contributions, one methodological and one physical.
We have built and validated a fully coupled framework for the structural and thermal evolution of rocky super-Earths, coupling the interior structure (\Zalmoxis), the mantle energetics (\Aragog), and the outgassed atmosphere (\atmodeller) within \PROTEUS, and we have introduced an oxygen-conserving redox treatment in which the planetary oxygen budget is conserved and the surface oxygen fugacity becomes a derived diagnostic.
With this framework we varied planet mass over $1$ to $10\,\Mearth$, the oxygen fugacity, the volatile inventory, the host star and irradiation, the core-mass fraction, and the initial thermal state, to establish how a rocky planet shrinks as it crystallises and what controls this phenomenon.

Investigating the coupled structural and thermal evolution of super-Earths, we find:

\begin{itemize}

\item The coupled \PROTEUS\ framework is a robust and validated tool for super-Earth evolution.
The interior-structure solver reproduces analytic limiting cases and published mass-radius and structure models up to $20\,\Mearth$, the mantle-energetics solver matches SPIDER for the magma ocean physics, and the full framework reproduces the CHILI magma ocean intercomparison (Section~\ref{sec:framework_validation}; Appendix~\ref{sec:validation}); the entropy-based energetics and the regularised phase boundaries keep the coupled integration stable across crystallisation.
The oxygen-conserving redox mode extends outgassing from an imposed surface fugacity to a conserved planetary oxygen inventory, recovering the fixed-fugacity result where that is physical.

\item The silicate mantle contracts measurably as it cools and crystallises, by about $11\,\%$ of the interior radius at $1\,\Mearth$ falling to about $9.5\,\%$ at $10\,\Mearth$.
The contraction is produced by the thinning silicate shell while the core radius stays nearly fixed, and it scales with the silicate mass fraction, so it is a property of the cooling mantle rather than of the bulk planet.

\item This contraction is robust across the grid.
It is set by the interior and its iron fraction, insensitive to the host star and the irradiation, with the solidified radius fixed by mass and composition.
A measured radius excess above the cold mass-radius relation therefore constrains the crystallisation state of the mantle, complementing the compositional information encoded in the bulk density.

\item Volatile-rich super-Earths and sub-Neptunes at and above about $5\,\Mearth$ may not solidify, and so may not realise this contraction.
Their thick outgassed atmospheres throttle the surface heat loss until the interior settles into a deep magma ocean at radiative equilibrium, limiting the contraction to about $3.5$ to $6.5\,\%$ and leaving the planet inflated; the volatile inventory thus decides whether the crystallisation contraction is realised.
The timescale on which such an interior would eventually solidify depends sensitively on the volatile budget and the atmosphere-loss history and is not derived here.

\item The treatment of oxidation state, not only its value, controls the outgassed atmosphere.
The speciation shifts from hydrogen and carbon monoxide through carbon dioxide as the fugacity rises, with sulfur shifting from hydrogen sulfide to sulfur dioxide, and the conserved free-oxygen budget increases monotonically; molecular oxygen comes to dominate only in the oxygen-conserving treatment at the oxidising extreme.
The fixed-fugacity prescription breaks down at the oxidising extreme, where it demands an atmosphere more massive than the planet, whereas the oxygen-conserving treatment stays well posed and is the appropriate tool for the corner relevant to abiotic oxygen.

\end{itemize}

In conclusion, the crystallisation contraction is a stable and interpretable signature of a cooling rocky interior, resolved here within a single coupled framework that treats interior structure, energetics, and outgassing on equal footing: a super-Earth caught mid-solidification shows a radius excess that records how much of its mantle remains molten, set by the mantle itself and largely independent of its surroundings.
We anticipate that coupling interior structure, energetics, and outgassing chemistry in this way will sharpen the interpretation of super-Earth radii and atmospheres as PLATO and the Roman Space Telescope enlarge the observed sample and JWST, and later the ELTs and LIFE, characterise its atmospheres.

\begin{acknowledgments}
The authors thank Johanna Teske for comments that improved the manuscript.
This research was supported by the Branco Weiss Foundation, the European Research Council (ERC) under the European Union's Horizon Europe research and innovation programme (MagmaWorlds, 101219807), the Alfred P.\ Sloan Foundation (AEThER, G-2025-25284), NASA's Nexus for Exoplanet System Science research coordination network (Alien Earths, 80NSSC21K0593), and the NWO NWA-ORC PRELIFE Consortium (NWA.1630.23.013).
M.A.\ is supported by the Swiss National Science Foundation through the Postdoc.Mobility fellowship, grant number 230229.
D.J.B.\ and P.A.S.\ acknowledge support from the Swiss State Secretariat for Education, Research and Innovation (SERI) under contract number MB22.00033 (ERC Starting Grant 2ATMO).
K.H.\ acknowledges support from the Belgian Science Policy Office (BELSPO) STELLA project Prf-2021-022, the Research Foundation Flanders (FWO) grant G014425N, and COST Action CA22133 (PLANETS).
H.N.\ acknowledges support from STFC grant UKRI1184.
R.C.\ acknowledges support from STFC grant ST/Y509139/1.
We thank the Center for Information Technology of the University of Groningen for providing access to the H\'abr\'ok high performance computing cluster.
\end{acknowledgments}

\vspace{5mm}
\noindent\textit{Data availability.}
The data and scripts to reproduce the figures in this manuscript are openly archived on Zenodo (\zdoi{22663462}) and maintained on GitHub (\href{https://github.com/FormingWorlds/superearth-interiors-figures}{FormingWorlds/superearth-interiors-figures}).

\vspace{5mm}
\software{
    \PROTEUS\ (v26.07.14, \zdoi{21358381}),
    \Aragog\ (v26.07.04, \zdoi{21196696}),
    \Zalmoxis\ (v26.07.13, \zdoi{21342353}),
    \PALEOS\ \citep{Attia2026subm},
    \CALLIOPE\ (v26.07.03, \zdoi{21162734}),
    \atmodeller\ (v1.0.1; \citealp{Bower2025ApJ99559B}),
    \AGNI\ (v1.9.4, \zdoi{15386789}; \citealp{Nicholls2025JOSS107726N}),
    MORS (v26.07.12, \zdoi{21315171}),
    ZEPHYRUS (v26.07.10, \zdoi{21301993}),
    Astropy \citep{AstropyCollaboration2018AJ156123A},
    NumPy \citep{Harris2020Natur585357H},
    SciPy \citep{Virtanen2020NatMe17261V},
    Matplotlib \citep{Hunter2007CSE990H},
    pandas \citep{McKinney2010SciPy56M}.
    This work made use of the Claude Code command-line tool \citep{AnthropicClaudeCode2026} for code assistance and language revision.
}

\bibliography{references}
\bibliographystyle{aasjournalv7}

\clearpage
\appendix
\restartappendixnumbering
\section{Model validation}
\label{sec:validation}
\suppressfloats[t]

\subsection{\Zalmoxis: analytic and limiting-case tests}
\label{sec:app:zalmoxis}

We verify the numerical machinery of \Zalmoxis\ against density distributions whose interior structure has an exact closed-form solution.
Each test prescribes a density law and compares the integrated mass, gravity, and pressure against the analytic result, so the checks are independent of the tabulated equation of state used in the coupled runs.
Most inject the density law directly into the structure integrator at a prescribed central pressure, isolating the differential-equation solver, the layer assignment, and the pressure-density coupling from any thermodynamic-table error; one instead registers the analytic law as a regular equation-of-state option and runs the complete solver, so that the outer mass-radius search, the Picard density iteration, and the Brent central-pressure root-find are verified in full.
The tabulated equation of state of the coupled runs is validated separately against published structure models in Appendix~\ref{sec:val:zalmoxis}, and the figure scripts are released with the manuscript data record.

\begin{figure*}
\centering
\includegraphics[width=\textwidth]{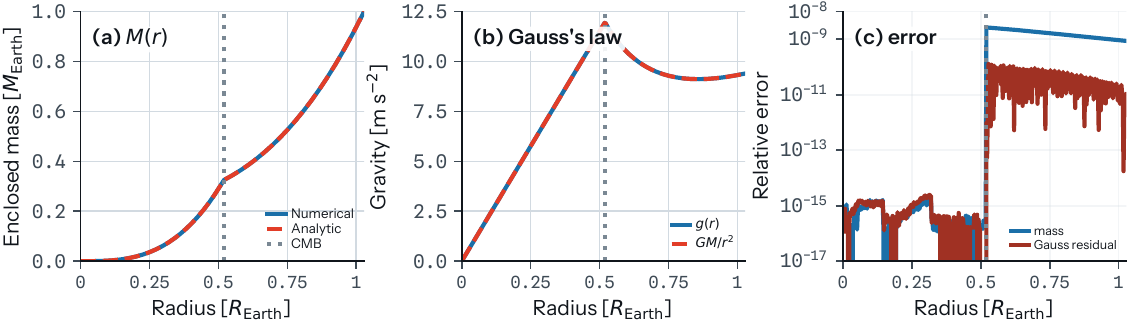}
\caption{
Two-layer constant-density sphere through the \Zalmoxis\ integrator, radius axis shared ($\rho_\mathrm{c} = 13{,}000$, $\rho_\mathrm{m} = 4000\,\mathrm{kg\,m^{-3}}$, core-mass fraction $0.325$); numerical solution solid, exact closed form dashed.
\textbf{(a)} Enclosed mass, \textbf{(b)} gravity against Gauss's law $g = GM/r^2$, and \textbf{(c)} the relative error of the mass and of the Gauss's-law residual; the dotted line marks the core-mantle boundary, with the residual at double-precision rounding ($\sim 10^{-15}$) through the core and a few parts in $10^{9}$ in the mantle.
}
\label{fig:val_zalmoxis_spheres}
\end{figure*}

A two-layer constant-density sphere tests the bare integrator across a density discontinuity (Figure~\ref{fig:val_zalmoxis_spheres}): with an iron-like core ($\rho_\mathrm{c} = 13{,}000\,\mathrm{kg\,m^{-3}}$) and a silicate-like mantle ($\rho_\mathrm{m} = 4000\,\mathrm{kg\,m^{-3}}$) at a core-mass fraction of $0.325$, the enclosed mass and gravity match the piecewise closed-form solution and Gauss's law $g = GM/r^2$, with the residual at double-precision rounding ($\sim 10^{-15}$) through the core, where the mass, gravity, and the $2g/r$ singularity at the centre are all exercised, and a few parts in $10^{9}$ in the mantle.

\begin{figure*}
\centering
\includegraphics[width=\textwidth]{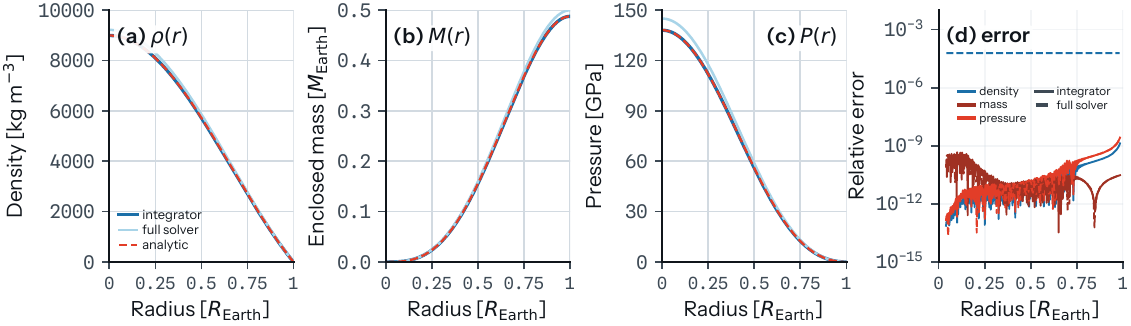}
\caption{
Polytrope of index $n = 1$ ($P = K\rho^2$, surface at $1\,\Rearth$) against the Lane-Emden solution $\rho = \rho_\mathrm{c}\sin\xi/\xi$, radius axis shared; columns are \textbf{(a)} density, \textbf{(b)} enclosed mass, \textbf{(c)} pressure, and \textbf{(d)} the relative error.
Each of (a-c) overlays the integrator (blue) at the analytic central pressure ($\rho_\mathrm{c} = 9000\,\mathrm{kg\,m^{-3}}$) and the complete solver (light blue) run with only the target mass prescribed, both against the analytic reference (red dashed).
The integrator matches to a few parts in $10^{9}$; the full solver converges to mass $0.004\,\%$, radius $0.14\,\%$ ($R_\mathrm{num}/R_\mathrm{exact} = 1.0014$), the interior mass and pressure profiles to a few parts in $10^{9}$ to $10^{10}$, and the density to the Picard iteration tolerance.
In (d), colour denotes the quantity (density, mass, pressure) and line style the method (solid integrator, dashed full solver).
}
\label{fig:val_zalmoxis_laneemden}
\end{figure*}

The constant-density tests leave the density fixed; the polytrope of index $n = 1$ closes this gap, because its equation of state $P = K\rho^2$ couples the density to the pressure the solver itself computes while still admitting an exact solution: $\rho(r) = \rho_\mathrm{c}\,\sin\xi/\xi$ with $\xi = r/\alpha$, $\alpha = \sqrt{K/2\pi G}$, and the surface at $\xi = \pi$, so $R = \pi\alpha$ is set by $K$ alone and we choose $K$ to place the surface at one Earth radius.
We run it two ways (Figure~\ref{fig:val_zalmoxis_laneemden}).
Injected into the integrator at the analytic central pressure, it reproduces the density, mass, and pressure to a few parts in $10^{9}$.
Registered as a regular equation-of-state option and driven through the complete solver with only the target mass prescribed, it converges to the same exact solution at the default tolerances: mass to $0.004\,\%$, radius to $0.14\,\%$, the interior mass and pressure profiles to a few parts in $10^{9}$ to $10^{10}$, and the density profile to the Picard iteration tolerance.
This is the only first-principles test that exercises the complete solver stack, not just the integrator, against an exact solution; it uses a single-composition body, so the multi-layer tabulated path is exercised instead against the literature models in Appendix~\ref{sec:val:zalmoxis}.

\begin{figure*}
\centering
\includegraphics[width=\textwidth]{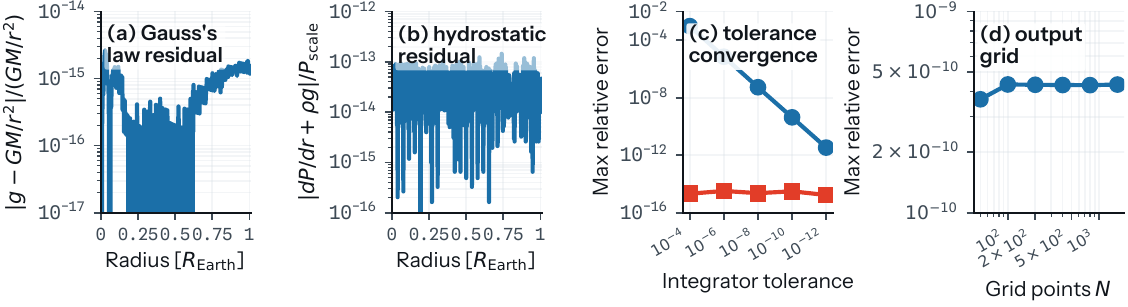}
\caption{
Conservation residuals and integrator convergence for the constant-density sphere and the $n = 1$ polytrope.
\textbf{(a)} Gauss's-law residual $|g - GM/r^2|/(GM/r^2)$, with $M(r)$ and $g(r)$ integrated independently, and \textbf{(b)} hydrostatic-balance residual $|\mathrm{d}P/\mathrm{d}r + \rho g|$ normalised by the central pressure scale, both at double-precision rounding and finite-difference truncation throughout.
\textbf{(c)} Maximum relative error against the integrator tolerance for the $n = 1$ polytrope (blue circles, non-polynomial) and the constant-density sphere (red squares, integrated exactly), and \textbf{(d)} against the output grid at fixed tolerance ($\mathrm{rtol} = 10^{-10}$): the polytrope tracks the tolerance to machine precision while the sphere stays at rounding, and the grid does not set the accuracy.
}
\label{fig:val_zalmoxis_diagnostics}
\end{figure*}

Two further diagnostics confirm that the converged solutions satisfy the conservation laws and converge under refinement (Figure~\ref{fig:val_zalmoxis_diagnostics}).
Panels (a, b) evaluate Gauss's law and the hydrostatic balance $\mathrm{d}P/\mathrm{d}r + \rho g = 0$ pointwise on a uniform sphere ($\rho = 5000\,\mathrm{kg\,m^{-3}}$); both residuals remain at double-precision rounding and second-order finite-difference truncation, with no systematic drift.
Panels (c, d) characterise the adaptive integrator: because the radial steps are set by the integrator tolerance rather than the output grid, the meaningful study is the error against tolerance, and the $n = 1$ polytrope, whose profile is not a polynomial, tracks the tolerance to machine precision, while the constant-density sphere, a low-degree polynomial integrated exactly, stays at rounding at every tolerance and every grid.

\subsection{\Zalmoxis: comparison with literature super-Earth structure models}
\label{sec:val:zalmoxis}
The tests of Appendix~\ref{sec:app:zalmoxis} verify the solver's numerics; here we give the full breakdown of the mass-radius comparison summarised in Section~\ref{sec:framework_validation} (Figure~\ref{fig:val_zalmoxis_mr}), validating the tabulated equation of state of the coupled runs against published structure models, together with the interior profiles.
We compute static mass-radius relations for bare, condensed planets with no volatile envelope or atmosphere, so that the radius is the solid surface of the core-mantle body and the comparison is made against the same quantity tabulated by the reference models.
The mass is swept from $0.5$ to $20\,\Mearth$ at four core-mass fractions spanning the plausible range, $0.01$ (rock), $0.325$ (Earth-like), $0.50$ (iron-rich), and $0.999$ (iron), with an iron core and an MgSiO$_3$ mantle held isothermal at $300\,\mathrm{K}$ to match the cold reference models.

Four independent references are used (Figure~\ref{fig:val_zalmoxis_mr}).
\Zalmoxis\ is run a second time with the built-in \citet{Seager2007ApJ6691279S} analytic equation of state, giving both an internal cross-check and a published benchmark.
MAGRATHEA \citep{Huang2022MNRAS5135256H} is an independent, standalone interior-structure code whose mineral-physics equations of state are distinct from the \PALEOS\ tables, run here in the matched two-layer configuration.
The \citet{Noack2020AA638A129N} analytic scaling relations, calibrated to full structure models over $0.8$-$2\,\Mearth$, provide a low-mass anchor, and the PREM-based tables of \citet{Zeng2016ApJ819127Z} furnish the residual baseline.

For rock and Earth-like compositions the \Zalmoxis\ radii agree with \citet{Zeng2016ApJ819127Z} to better than $1.3\,\%$ across the full mass range and with MAGRATHEA to better than $3\,\%$, well within the spread among the reference models.
The pure-iron endmember is the only case exceeding $5\,\%$ against \citet{Zeng2016ApJ819127Z}: PALEOS lies $\sim 5\,\%$ below the PREM-extrapolated iron curve and MAGRATHEA about $4$ to $7\,\%$ below, the offset largest at low mass, and the three equations of state agree with one another to $1.6\,\%$ below $2\,\Mearth$, widening to $\sim 4\,\%$ by $20\,\Mearth$ as the \citet{Seager2007ApJ6691279S} iron equation of state stiffens.
The offset is consistent with a difference in the adopted iron equation of state at terapascal pressures rather than a solver error, though PALEOS and the \citet{Seager2007ApJ6691279S} equation of state share the \Zalmoxis\ integrator, so their mutual agreement is only a partial cross-check.
The \citet{Noack2020AA638A129N} scaling tracks the full solvers over its $0.8$ to $2\,\Mearth$ calibration range, the span shown in Figure~\ref{fig:val_zalmoxis_mr}a; above it the relation is an extrapolation, which is the regime where the coupled super-Earth evolution requires the full structure solver.

\begin{figure*}
\centering
\includegraphics[width=\textwidth]{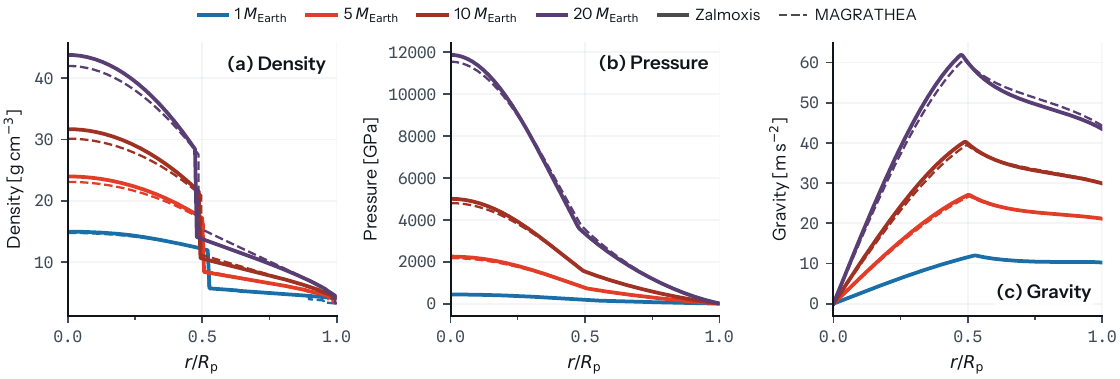}
\caption{
Interior profiles of \Zalmoxis\ (solid) and MAGRATHEA (dashed) for the Earth-like composition (core-mass fraction $0.325$) at $1$, $5$, $10$, and $20\,\Mearth$, against normalised radius.
\textbf{(a)} Density and \textbf{(b)} pressure rise toward the centre; \textbf{(c)} gravity peaks at the core-mantle boundary.
The dashed MAGRATHEA curves overlap with the solid \Zalmoxis\ curves over most of the radius and are visible mainly near the centre and at the core-mantle density step, which the two codes place at nearly the same radius fraction; the central values match to better than $1.5\,\%$ at $1\,\Mearth$ and to within $\sim 5\,\%$ over $10$ to $20\,\Mearth$, reflecting the high-pressure iron equation-of-state difference.
}
\label{fig:val_zalmoxis_profiles}
\end{figure*}

The interior profiles show the same pattern (Figure~\ref{fig:val_zalmoxis_profiles}).
At $1$, $5$, $10$, and $20\,\Mearth$ the density, pressure, and gravity profiles of \Zalmoxis\ and MAGRATHEA place the core-mantle density jump at nearly the same radius fraction (agreeing to within $\sim 3\,\%$); the central densities and pressures match to better than $1.5\,\%$ at $1\,\Mearth$ and differ by up to $\sim 5\,\%$ over $10$ to $20\,\Mearth$, the same high-pressure iron equation-of-state difference seen in the mass-radius residuals (running \Zalmoxis\ on a representative hot adiabat rather than the $300\,\mathrm{K}$ isotherm leaves the offset essentially unchanged, confirming it is not a thermal artifact).
The tabulated equation of state therefore reproduces the mass-radius relation and internal structure to within the few-per-cent spread among the comparison models for rock and Earth-like compositions, and to $\sim 5\,\%$ for the pure-iron endmember, where the equations of state themselves disagree most.

\subsection{\Aragog: heat-equation limiting-case tests}
\label{sec:app:aragog}
We verify the \Aragog\ energy solver against limiting cases of the heat equation for which the interior evolution has a closed-form solution.
\Aragog\ integrates the entropy-form energy balance $\rho T\,\partial S/\partial t = -r^{-2}\,\partial_r(r^2 F) + \rho H$ with a flux $F$ assembled from conduction, convection, gravitational separation, and compositional mixing.
Each test runs the solver in its constant-property mode: constant density, heat capacity, and thermal conductivity with the analytic temperature relation $T(S) = T_\mathrm{ref}\exp[(S - S_\mathrm{ref})/c_p]$ and no tabulated equation of state, in which the entropy balance reduces exactly to the classical temperature heat equation.
This is also the configuration used for the SPIDER cross-comparison (Appendix~\ref{sec:val:aragog}).
The same code path that integrates the coupled runs is exercised; only the material model is simplified so that the comparison is against an exact solution rather than a tabulated one.
The three tests probe complementary parts of the solver: the steady profiles verify the spatial operator and the boundary fluxes by holding an injected exact steady state, the conservation and invariance laws verify the global energy bookkeeping under evolution, and the transient eigenmode decay verifies the time integrator.

\begin{figure*}
\centering
\includegraphics[width=\textwidth]{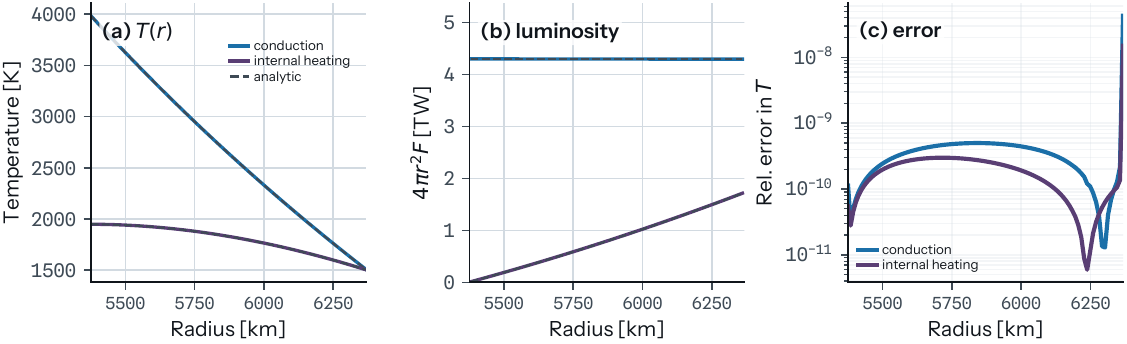}
\caption{
Steady conduction tests of the \Aragog\ energy solver in constant-property mode; the exact profile is imposed as the initial state and held by the solver in its default configuration.
Each panel overlays two cases against their analytic profiles (grey dashed): a shell carrying a steady conductive flux with no internal heating (conduction, the matching flux prescribed at both boundaries), and a uniform internal heating $\rho H$ with an insulated base (internal heating).
\textbf{(a)} Temperature, against the exact $T(r) = A/r + B$ for the conduction case and the exact internally-heated profile for the heated case, where $A$ and $B$ are integration constants fixed by the boundary temperatures and $H$ is the specific heating rate.
\textbf{(b)} The luminosity $4\pi r^2 F$, uniform for the conduction case and rising as $r^3 - R_\mathrm{c}^3$ toward the total heating power for the heated case ($R_\mathrm{c}$ the inner radius), shown on the interior nodes where the flux is assembled by the same operator that advances the solution (the boundary nodes carry the prescribed boundary flux and are excluded).
\textbf{(c)} The relative error in temperature for both cases.
}
\label{fig:val_aragog_conduction}
\end{figure*}

We first impose an exact steady conduction profile as the initial state and confirm that the solver holds it (Figure~\ref{fig:val_aragog_conduction}).
With conduction the only active transport and no internal heating, the steady solution of $\nabla\cdot(k\nabla T) = 0$ on the shell is $T(r) = A/r + B$, with integration constants $A$ and $B$ fixed by the boundary temperatures.
Supplied as the initial entropy and evolved under the matching conductive fluxes prescribed at both boundaries, it is held to a relative error below $5\times 10^{-8}$, and the conductive luminosity $4\pi r^2 F$ stays uniform across the interior to $0.11\,\%$ as the flux divergence vanishes.
The luminosity is evaluated on the interior nodes, where it is assembled by the same staggered-node operator that advances the solution; the two boundary nodes instead carry the prescribed boundary flux and are not directly comparable.
Imposing instead a uniform volumetric heating $\rho H$ with an insulated base, the matching internally-heated steady profile $T(r) = T_\mathrm{s} + (\rho H/3k)\,[R_\mathrm{c}^3(R_\mathrm{p}^{-1} - r^{-1}) + (R_\mathrm{p}^2 - r^2)/2]$ is again supplied as the initial state and held to better than $2\times 10^{-8}$, with the luminosity rising as $r^3 - R_\mathrm{c}^3$ and approaching the total heating power.

\begin{figure*}
\centering
\includegraphics[width=\textwidth]{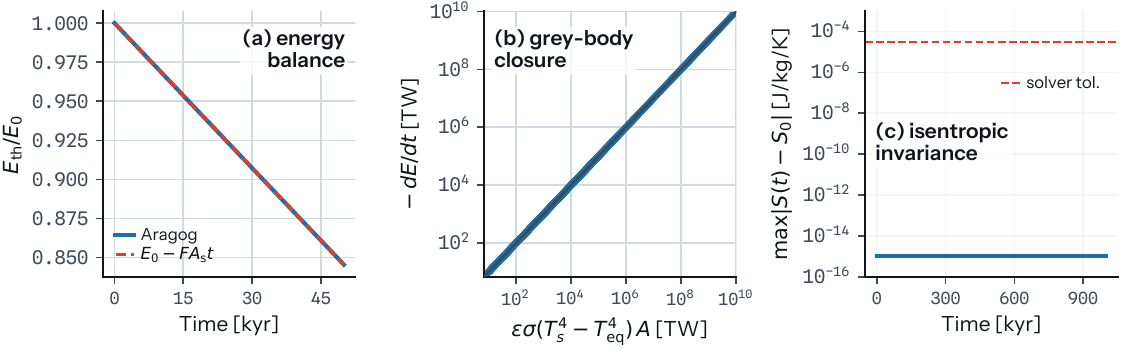}
\caption{
Conservation and invariance laws for the \Aragog\ energy solver, holding for any valid run.
$A_\mathrm{s}$ is the surface area, $\varepsilon$ the emissivity, $\sigma$ the Stefan-Boltzmann constant, $T_\mathrm{s}$ the surface temperature, and $T_\mathrm{eq}$ the equilibrium temperature.
\textbf{(a)} Thermal energy (solid blue) under a prescribed surface flux with an insulated base, against the exact linear decay $E_0 - F A_\mathrm{s} t$ (dashed red).
\textbf{(b)} Grey-body closure: the energy-loss rate $-\mathrm{d}E/\mathrm{d}t$ against the Stefan-Boltzmann surface power $\varepsilon\sigma(T_\mathrm{s}^4 - T_\mathrm{eq}^4)A_\mathrm{s}$ over the cooling, the grey line marking equality.
\textbf{(c)} Drift of a uniform isentropic state between insulated boundaries, far below the effective per-step integrator tolerance (dashed).
}
\label{fig:val_aragog_conservation}
\end{figure*}

A second set of tests checks the conservation and invariance laws that hold for any valid run (Figure~\ref{fig:val_aragog_conservation}).
Under a prescribed surface flux $F$ with an insulated base the thermal energy decreases linearly, $E(t) = E_0 - F A_\mathrm{s} t$ with $A_\mathrm{s}$ the surface area, which the solver reproduces with $\mathrm{d}E/\mathrm{d}t = -FA_\mathrm{s}$ to numerical precision.
Under a grey-body surface the radiative closure is reproduced: the energy-loss rate $-\mathrm{d}E/\mathrm{d}t$ tracks $\varepsilon\sigma(T_\mathrm{s}^4 - T_\mathrm{eq}^4)\,A_\mathrm{s}$ to better than $0.2\,\%$ across nearly nine decades of cooling.
A uniform isentropic state held between insulated boundaries is invariant to machine precision, many orders of magnitude below the integrator tolerance, confirming that the solver generates no spurious entropy from a uniform isentropic state.

\begin{figure*}
\centering
\includegraphics[width=\textwidth]{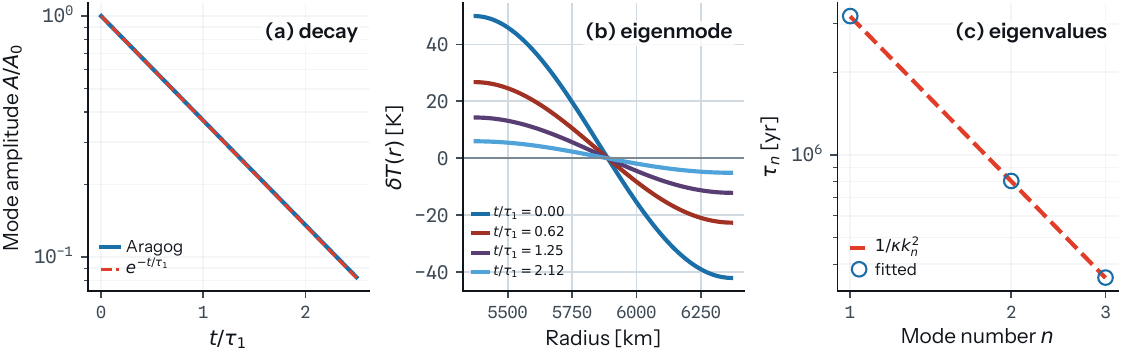}
\caption{
Transient conduction in an insulated shell, testing the \Aragog\ time integrator.
\textbf{(a)} Decay of the seeded fundamental eigenmode amplitude (solid blue) against $\exp(-t/\tau_1)$ (dashed red).
\textbf{(b)} The fundamental mode shape $\delta T(r)$ at successive times (coloured by $t/\tau_1$), preserved as the amplitude decays.
\textbf{(c)} Fitted decay timescales (circles) for the first three modes against the analytic Neumann eigenvalues $1/\kappa k_n^2$ (dashed red), with $\kappa$ the thermal diffusivity.
}
\label{fig:val_aragog_transient}
\end{figure*}

The time integrator is checked against the transient decay of conduction eigenmodes (Figure~\ref{fig:val_aragog_transient}).
For an insulated shell the perturbation eigenmodes $\delta T_n(r) \propto [v(R_\mathrm{c})\sin k_n r + u(R_\mathrm{c})\cos k_n r]/r$ decay as $\exp(-t/\tau_n)$ with $\tau_n = 1/(\kappa k_n^2)$ and thermal diffusivity $\kappa = k/(\rho c_p)$, where $u(r) = k_n r\cos k_n r - \sin k_n r$ and $v(r) = k_n r\sin k_n r + \cos k_n r$ are the radial combinations set by the zero-flux condition, and the wavenumbers $k_n$ solve the Neumann dispersion relation $v(R_\mathrm{c})\,u(R_\mathrm{p}) = v(R_\mathrm{p})\,u(R_\mathrm{c})$ for the spherical shell.
The thermal conductivity is raised for this test so that the diffusion time $\tau_n = 1/(\kappa k_n^2)$ is short enough to integrate over several decay times; because the analytic eigenvalues scale with the same $\kappa$, the comparison is unaffected, and the solver, the temperature relation, and the time integrator are otherwise those of the coupled runs.
Seeding each of the first three modes in turn, the fitted decay timescales match the analytic eigenvalues to better than $0.3\,\%$, the largest residual being $0.29\,\%$ for the third mode.

\subsection{\Aragog: comparison with SPIDER}
\label{sec:val:aragog}
\Aragog\ is the entropy-method successor to SPIDER \citep{Bower2018PEPI27449B,Bower2019AA631A103B,Bower2022PSJ393B} and was written to reproduce its magma ocean physics (Section~\ref{sec:methods:aragog}).
We cross-validate the two mantle solvers in two complementary limits: a constant-property limit in which the governing equation has a closed form, and a realistic magma ocean state evaluated with the tabulated equation of state of the coupled runs.
Both comparisons exercise the as-used code path of each solver on identical inputs, so any difference is attributable to the solver itself and not to the configuration.

We first compare the two solvers in the constant-property heat-equation limit of Appendix~\ref{sec:app:aragog} (Figure~\ref{fig:val_aragog_spider_const}).
A linear entropy profile is imposed in an insulated mantle shell and relaxed by pure conduction, with the thermal conductivity raised so that the diffusion time is short enough to integrate over; in this limit both solvers reduce to the classical heat equation with the analytic temperature relation $T(S) = T_\mathrm{ref}\exp[(S - S_\mathrm{ref})/c_p]$.
SPIDER reproduces this analytic relation to better than $10^{-5}$, confirming that it integrates the same reduced equation.
Supplied with the identical initial profile, the two solvers track the same homogenisation: over most of the interior the temperature profiles agree to better than $1\,\%$, and at $50\,\mathrm{Myr}$, about two conductive diffusion times of the shell, the difference is $0.3\,\%$ at the surface and grows to $3\,\%$ at the insulated core.
The residual is localised at the core and accumulates with time: over the insulated run SPIDER's total energy is not exactly conserved but rises by $0.83\,\%$, whereas \Aragog\ conserves it to $0.05\,\%$, and the core-temperature offset closely tracks this energy gain (correlation coefficient $0.92$).
The difference is therefore numerical, in the discretisation of the insulated boundary, rather than a physical disagreement, as both solvers reproduce the analytic temperature-entropy relation at the node level; this dynamic test exercises the conduction operator and the time integrator only, the convective transport being compared separately below.

\begin{figure*}
\centering
\includegraphics[width=\textwidth]{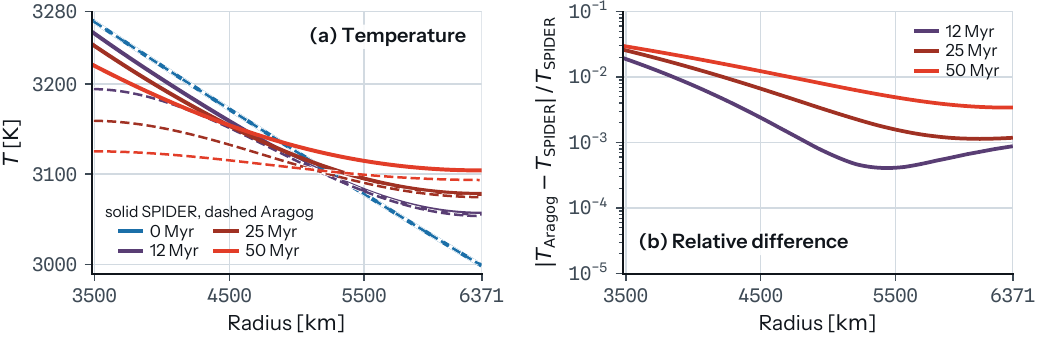}
\caption{
Cross-validation of the \Aragog\ and SPIDER energy solvers in the constant-property heat-equation limit: a linear entropy profile in an insulated mantle shell relaxed by pure conduction (raised conductivity so the diffusion time is integrable).
\textbf{(a)} Temperature profiles at four times from the initial state to $50\,\mathrm{Myr}$ (one colour per time, given in the legend): SPIDER (solid) and \Aragog\ (dashed) homogenise the steep initial profile together.
\textbf{(b)} Radius-resolved relative difference between the two solvers at the later times, growing from a few tenths of a per cent in the interior to about $3\,\%$ at the insulated core ($3480\,\mathrm{km}$) by $50\,\mathrm{Myr}$.
The radius runs from the core-mantle boundary ($3480\,\mathrm{km}$) to the surface ($6371\,\mathrm{km}$).
}
\label{fig:val_aragog_spider_const}
\end{figure*}

\begin{figure*}
\centering
\includegraphics[width=\textwidth]{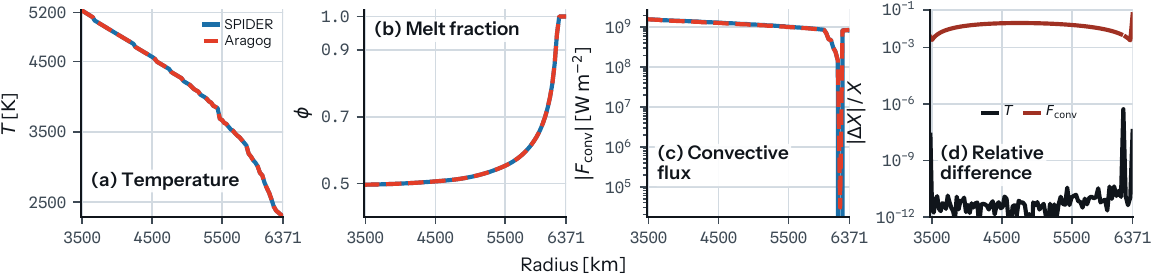}
\caption{
Cross-validation of the \Aragog\ and SPIDER energy solvers on a realistic magma ocean state with the PALEOS MgSiO$_3$ equation of state; both solvers read bit-identical pressure-entropy tables on the identical Adams-Williamson mesh and are evaluated on the same entropy profile.
\textbf{(a)} Temperature and \textbf{(b)} melt fraction $\phi$ against radius for SPIDER (solid) and \Aragog\ (dashed); the curves overlie to within $10^{-6}$ because both solvers interpolate the same tables.
\textbf{(c)} Convective heat flux $|F_\mathrm{conv}|$ on a logarithmic scale, near-overlying across the convecting mantle.
\textbf{(d)} Relative difference between the two solvers in temperature (near machine precision, below $10^{-6}$ across the interior) and in convective flux (about $1.5\,\%$ across the convecting interior, rising to about $5.7\,\%$ at the fully molten surface node).
The radius runs from the core-mantle boundary ($3480\,\mathrm{km}$) to the surface ($6371\,\mathrm{km}$).
}
\label{fig:val_aragog_spider_static}
\end{figure*}

We then compare the two solvers on a realistic magma ocean state with the \PALEOS\ MgSiO$_3$ equation of state of the coupled runs (Figure~\ref{fig:val_aragog_spider_static}).
Both solvers read bit-identical pressure-entropy tables on the identical Adams-Williamson mesh and are evaluated on the same entropy profile, a single representative partially crystallised magma ocean in which the mantle convects throughout and the melt fraction rises from $0.50$ at the base to unity at the surface.
Because both solvers interpolate the same tables on the same mesh, the thermodynamic fields they compute, the temperature, the melt fraction, and the density, agree to within $10^{-6}$; this confirms that the equation-of-state and phase-boundary lookups are implemented consistently between the codes rather than testing them independently.
The convective heat flux is the one field that genuinely exercises the mixing-length transport and the mush-region viscosity blending: it agrees to about $1.5\,\%$ across the convecting interior, rising to about $5.7\,\%$ at the fully molten surface node, where the flux is most sensitive to the boundary discretisation; the residual reflects the differing eddy-diffusivity formulation in the two-phase region.
No test in this appendix evolves convective transport dynamically, the constant-property comparison being conduction-only and the present comparison an instantaneous state; the fully coupled, time-dependent trajectory through the mush region, including the atmosphere and the long-term solidification, is assessed separately in the coupled CHILI Earth comparison (Appendix~\ref{sec:val:proteus}).

\subsection{\PROTEUS: the CHILI magma ocean intercomparison}
\label{sec:val:proteus}
The preceding subsections isolate individual modules; this one exercises the fully coupled framework, detailing the intercomparison summarised in Section~\ref{sec:framework_validation} (Figure~\ref{fig:val_chili}).
We run \PROTEUS\ in its full coupled configuration, the \Zalmoxis\ interior structure, the \Aragog\ interior energetics, the AGNI radiative-convective atmosphere, and the equilibrium outgassing module, on the community CHILI magma ocean intercomparison protocol \citep{Lichtenberg2026PSJ7108L} and compare the result against the six other coupled atmosphere-interior models that submitted to it, GOOEY, NEONGOOEY, PACMAN, LINCS, MOAI, and PlanAtMO, as well as the earlier \PROTEUS\ submission, which used SPIDER and an Adams-Williamson interior structure.
The submitted intercomparison results are drawn from \citet{Nicholls2026arXiv260624757N}.
The protocol fixes the inputs that all models share: a volatile budget of three Earth oceans of hydrogen ($4.7\times10^{20}\,\mathrm{kg}$) and $2.73\times10^{20}\,\mathrm{kg}$ of carbon, free to cycle between the mantle and the atmosphere, an oxygen fugacity buffered to IW+4, a Bond albedo of $0.1$, an inert core at a fixed radius fraction, and a fully molten start at a stellar age of $50\,\mathrm{Myr}$.
Each model evolves the magma ocean to its own termination criterion, and the runs are compared at common melt-fraction milestones down to $5\,\%$.
Two cases are run, a Nominal Earth ($1\,\Mearth$ at $1\,\mathrm{AU}$) and a Nominal Venus ($0.815\,\Mearth$ at $0.723\,\mathrm{AU}$).
This is the dynamic, fully coupled evolution through the mush region that Appendix~\ref{sec:val:aragog} deferred here.

The global melt fraction tracks the magma ocean from its molten start through solidification (Figure~\ref{fig:val_chili}a).
For Nominal Earth the current \PROTEUS\ run reaches $5\,\%$ melt at $1.34\,\mathrm{Myr}$, within the ensemble spread; the intercomparison finds every Earth model solidifies within about four million years, consistent with empirical constraints on Earth's early history.
\PROTEUS\ lies on the slow side of that spread, which ranges from about $0.01$ to $2\,\mathrm{Myr}$ across the other models and reflects genuine differences in atmospheric opacity, volatile partitioning, and interior convection rather than numerical disagreement on a shared problem; the intercomparison attributes \PROTEUS's slow early cooling to its treatment of mantle dynamics and its melting curves.
Venus is more diverse: the current \PROTEUS\ run cools continuously, reaching about $5\,\%$ melt by $2.2\,\mathrm{Myr}$, whereas several models, including the earlier \PROTEUS\ submission, instead enter an extended radiative-equilibrium plateau that sustains the magma ocean for up to $\sim50\,\mathrm{Myr}$; this regime change between the two \PROTEUS\ runs follows from the change in interior modules rather than the shared protocol.
Both \PROTEUS\ configurations lie within the community ensemble.

The interior diagnostics over the same solidification track the ensemble as well (Figure~\ref{fig:val_chili}b,c).
The surface temperature falls from the molten state through the mush region, tracking the other models in shape while plateauing at the cool end of the ensemble, and the rheological front, the radius at which the mantle crosses the rheological transition from a convecting melt-dominated rheology to a solid-like one, advances outward from the core-mantle boundary as crystallisation proceeds.
Like the other models, \PROTEUS\ cools rapidly at first and then stalls as the mantle crosses the rheological transition, where the stiffening mush region sharply reduces the convective heat flux.
The front advances toward a radius larger than the present-day solid Earth because \Zalmoxis\ computes the interior structure once, at the hot molten start: the planet radius then reflects the thermal expansion of the silicate melt and is not recontracted as the mantle cools, a reminder that accounting for melt thermal expansion materially changes the inferred radius; this one-time structure update offsets the absolute front radius without affecting the melt-fraction and surface-temperature tracks on which the ensemble comparison rests.
The agreement of the coupled surface-temperature and solidification-front evolution with the community ensemble, on top of the module-level validations above, indicates that the coupled framework advances the magma ocean from melt through solidification consistently with the independent coupled models in the intercomparison.

\subsection{\CALLIOPE\ vs \atmodeller: cross-module thermochemical comparison}
\label{sec:app:calliope_atmodeller}
\PROTEUS\ includes two outgassing modules, \CALLIOPE\ and \atmodeller; the runs of this work use \atmodeller\ in oxygen-conserving mode, with \CALLIOPE\ retained as an independent cross-validation reference (Section~\ref{sec:methods:atmodeller}).
The module agreement is summarised in Section~\ref{sec:framework_validation} (Figure~\ref{fig:val_calliope_atmodeller}); here we give the full comparison, including the equilibrium speciation.
Both modules invert the same closure through a shared oxygen-conserving entry point: the supplied oxygen budget is the volatile oxygen that participates in atmospheric and dissolved chemistry, and the iron-w\"ustite offset $\Delta\mathrm{IW}$ is solved as an additional unknown.
They differ in their oxygen-fugacity buffer, their selection of solubility laws, their gas-phase equation of state, their equilibrium-constant fits, and their solver architecture.
To isolate those internal differences from the inputs, both are run at one shared state: the Earth bulk-silicate inventory of \citet{Krijt2023ASPC5341031K} (hydrogen, carbon, nitrogen, and sulfur summed across mantle and atmosphere), with the volatile-oxygen reference fixed by a \CALLIOPE\ buffered-mode call at the \citet{Sossi2020SciA61387S} upper-mantle anchor $\Delta\mathrm{IW} = +3.5$, melt fraction unity, and magma temperatures from $1800$ to $3000\,\mathrm{K}$.
The shared solubility laws are matched by construction, water on the peridotite calibration of \citet{Sossi2023EPSL60117894S}, carbon dioxide on \citet{DIXON1995JPet361607D}, and nitrogen on \citet{Dasgupta2022GeCoA336291D}; the sulfur law, the equilibrium constants, and the minor-species solubilities are not.

The dominant cross-module systematic is the oxygen-fugacity buffer.
\CALLIOPE's default buffer \citep{Fischer2011EPSL304496F} lies within about $0.2\,\mathrm{dex}$ of \atmodeller's Hirschmann composite \citep{ONeill1993CoMP114296O,Hirschmann2021GeCoA31374H} across the magma ocean range, whereas the legacy option \citep{ONeill2002ChGeo186151O} diverges from it by up to about $1\,\mathrm{dex}$ at the hottest end.
The converged offsets follow this expectation (Figure~\ref{fig:val_calliope_atmodeller}a): the \atmodeller\ curve tracks \CALLIOPE\ with the Fischer buffer to within a few tenths of a dex, and the curve obtained by shifting \CALLIOPE-Fischer by the analytic Hirschmann-minus-Fischer buffer offset hugs \atmodeller, so the remaining gap is set by the buffer convention rather than by the chemistry.

The buffer-corrected residual quantifies what is left (Figure~\ref{fig:val_calliope_atmodeller}b).
The raw gap under the Fischer default stays within $0.25\,\mathrm{dex}$ across the whole sweep and is $0.16\,\mathrm{dex}$ at $2000\,\mathrm{K}$, while the legacy gap grows from $0.07\,\mathrm{dex}$ at $1800\,\mathrm{K}$ to $1.23\,\mathrm{dex}$ at $3000\,\mathrm{K}$.
After the analytic buffer offset is removed, the residual chemistry-level disagreement is within $\pm0.1\,\mathrm{dex}$ below $2000\,\mathrm{K}$ and rises to about $0.28\,\mathrm{dex}$ toward the hottest, most oxidising end, where the sulfate-regime difference between the \citet{Gaillard2022EPSL57717255G} and \citet{Boulliung2023CoMP17856B} sulfur solubility laws is largest.
Both modules place Earth inside the empirical upper-mantle window of \citet{Frost2008AREPS36389F}, $\Delta\mathrm{IW}$ between $+1$ and $+5$, with \CALLIOPE\ at $+3.24$ and \atmodeller\ at $+3.08$ at $2000\,\mathrm{K}$, a raw default-configuration offset of $0.16\,\mathrm{dex}$, so the cross-module offset is small against the petrological uncertainty on Earth's mantle oxidation state.
The two outgassing modules therefore agree to a few tenths of a dex in derived oxygen fugacity across the magma ocean range; the buffer convention dominates the small residual at lower temperature and the sulfur solubility law dominates it toward the hottest, most oxidising end.

\begin{figure*}
\centering
\includegraphics[width=\textwidth]{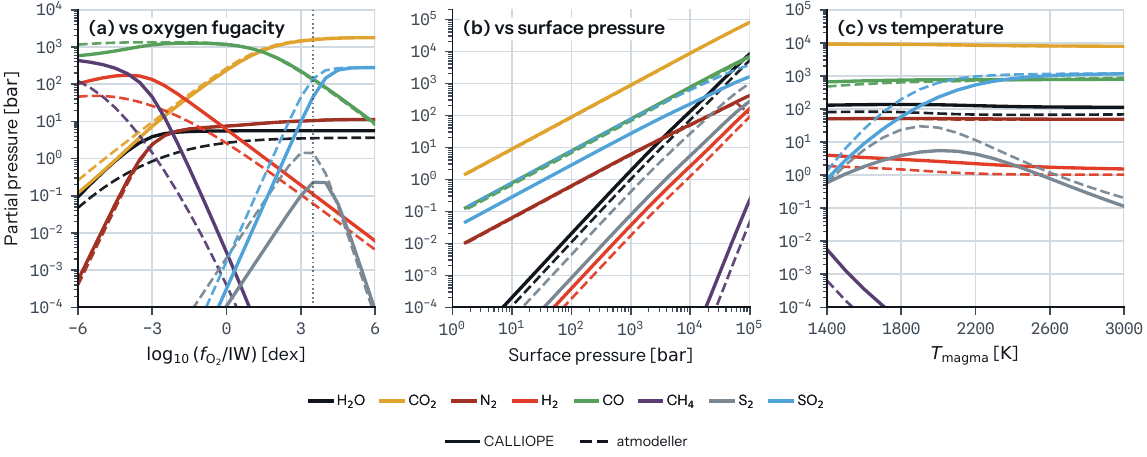}
\caption{
Equilibrium gas speciation of the two outgassing modules in fixed-fugacity mode at the Earth bulk-silicate inventory \citep{Krijt2023ASPC5341031K}, melt fraction unity; \CALLIOPE\ solid, \atmodeller\ dashed, one colour per species.
\textbf{(a)} Partial pressures of the eight major C-H-N-S-O species against oxygen fugacity (the IW-buffer offset) at the fixed inventory and $2000\,\mathrm{K}$; the dotted vertical marks the IW$+3.5$ anchor used in panels (b) and (c).
\textbf{(b)} The same against total surface pressure, the volatile inventory scaled at fixed oxygen fugacity IW$+3.5$ and $2000\,\mathrm{K}$.
\textbf{(c)} The same against magma temperature, at fixed oxygen fugacity IW$+3.5$ and fixed total surface pressure $10^{4}\,\mathrm{bar}$, the inventory solved at each temperature so the surface pressure is held constant.
All panels use an ideal gas phase, so the high-pressure end of (b) is indicative rather than real-gas accurate.
}
\label{fig:val_calliope_atmodeller_speciation}
\end{figure*}

The same modules can be compared at the level of the full equilibrium speciation rather than the single derived oxygen fugacity (Figure~\ref{fig:val_calliope_atmodeller_speciation}).
Run in fixed-fugacity mode at the same Earth inventory and $2000\,\mathrm{K}$, both reproduce the standard sequence in which the reduced carriers (CO, H$_2$, CH$_4$) give way to the oxidised ones (CO$_2$, H$_2$O, SO$_2$) as the imposed oxygen fugacity increases (Figure~\ref{fig:val_calliope_atmodeller_speciation}a); the comparison persists when the inventory is scaled at fixed oxygen fugacity, from thin atmospheres up to surface pressures of order $10^{5}\,\mathrm{bar}$ (Figure~\ref{fig:val_calliope_atmodeller_speciation}b), and when temperature is varied at fixed oxygen fugacity and surface pressure, where both reproduce the expected trends of a rising CO/CO$_2$ ratio and a falling CH$_4$ abundance with temperature (Figure~\ref{fig:val_calliope_atmodeller_speciation}c).
The systematic differences are concentrated in two channels.
Sulfur is the larger: \atmodeller's S$_2$ and SO$_2$ partial pressures exceed \CALLIOPE's, S$_2$ by up to about a factor of twenty and SO$_2$ by about five over most of the oxygen-fugacity range, because \atmodeller\ adopts the sulfate solubility of \citet{Boulliung2023CoMP17856B} where \CALLIOPE\ uses the sulfide-only law of \citet{Gaillard2022EPSL57717255G}; the two laws agree only toward the most oxidising end, and this surplus sulfur is what raises \atmodeller's total surface pressure, for example $1910$ against $1755\,\mathrm{bar}$ at IW$+3.5$.
The reduced carbon and hydrogen carriers are the second: \CALLIOPE\ treats H$_2$, CO, and CH$_4$ as insoluble \citep{Bower2022PSJ393B} whereas \atmodeller\ assigns them basalt solubility laws \citep{Hirschmann2012EPSL34148H,Yoshioka2019GeCoA259129Y,Ardia2013GeCoA11452A}, so the two diverge at reducing conditions where these species dominate the atmosphere: \atmodeller\ holds back some H$_2$ and CH$_4$ in the melt, while the reduced-carbon budget redistributes toward a higher CO partial pressure.
Both are the speciation-level expression of the same solubility-law choices that set the residual in the oxygen-fugacity comparison above.

The grid runs adopt \atmodeller\ throughout, so these differences do not propagate into the coupled evolution, and \CALLIOPE\ supplies an independent check on the volatile partitioning.

\subsection{Conserved oxygen budget across the oxygen-fugacity sweep}
\label{sec:app:oxygen_budget}
The oxygen-conserving twins of Section~\ref{sec:results:redox_mode} each conserve the total volatile-system oxygen set by their fixed-fugacity partner at the initial outgassing equilibrium.
Table~\ref{tab:oxygen_budget} lists that conserved oxygen mass for the fiducial $5\,\Mearth$ reference planet at every level of the imposed iron-w\"ustite offset, together with its ratio to the fixed hydrogen, carbon, nitrogen, and sulfur inventory.
The oxygen budget rises by more than two orders of magnitude across the sweep, from comparable to the summed C-H-N-S mass at the reducing end to nearly $380$ times it at the most oxidising level retained, and this monotonic growth is what makes the oxygen-conserving closure single-valued.

\begin{deluxetable}{lcc}
\tablecaption{Conserved volatile-system oxygen budget of the fiducial $5\,\Mearth$ reference planet across the imposed oxygen-fugacity sweep.\label{tab:oxygen_budget}}
\tabletypesize{\footnotesize}
\tablehead{\colhead{$\Delta\mathrm{IW}$} & \colhead{$M_\mathrm{O}$ (kg)} & \colhead{$M_\mathrm{O}/M_\mathrm{CHNS}$}}
\startdata
$-6$ & $4.28\times10^{21}$ & $1.07$ \\
$-5$ & $5.97\times10^{21}$ & $1.50$ \\
$-4$ & $8.26\times10^{21}$ & $2.07$ \\
$-3$ & $1.10\times10^{22}$ & $2.76$ \\
$-2$ & $1.38\times10^{22}$ & $3.45$ \\
$-1$ & $1.61\times10^{22}$ & $4.05$ \\
$0$ & $1.80\times10^{22}$ & $4.52$ \\
$+1$ & $1.96\times10^{22}$ & $4.92$ \\
$+2$ & $2.16\times10^{22}$ & $5.41$ \\
$+3$ & $2.94\times10^{22}$ & $7.38$ \\
$+4$ & $1.08\times10^{23}$ & $27.1$ \\
$+5$ & $1.50\times10^{24}$ & $377$ \\
\enddata
\tablecomments{The oxygen mass $M_\mathrm{O}$ is the volatile-bound oxygen, in the atmosphere and dissolved in the melt, and excludes the silicate mantle oxides; it is set by the fixed-fugacity twin at the initial outgassing equilibrium and then conserved. The hydrogen, carbon, nitrogen, and sulfur inventory is held fixed across the sweep at $M_\mathrm{H}=2.35\times10^{21}$, $M_\mathrm{C}=1.365\times10^{21}$, $M_\mathrm{N}=2.5\times10^{19}$, and $M_\mathrm{S}=2.5\times10^{20}\,\mathrm{kg}$, summing to $M_\mathrm{CHNS}=3.99\times10^{21}\,\mathrm{kg}$.}
\end{deluxetable}

\end{document}